\documentclass[12pt, a4paper]{article}

\pdfoutput=1

\usepackage{amsmath}
\usepackage{amsfonts}
\usepackage{amssymb}
\usepackage{latexsym}
\usepackage{graphicx}
\usepackage{color}
\usepackage{mathrsfs}
\usepackage{slashed}
\usepackage{soul}

\usepackage{cite}
\usepackage{hyperref}

\usepackage{enumerate}
\usepackage{multirow}
\usepackage[table]{xcolor}
\usepackage{colortbl}

\usepackage{float}

\usepackage{comment}
\includecomment{toexclude} 
\excludecomment{addsections} 

\numberwithin{equation}{section}

\hypersetup{colorlinks, citecolor=bluscuro, linkcolor=magenta, urlcolor=bluscuro}
\definecolor{rossos}{rgb}{0.7,0,0.3}
\definecolor{violachiaro}{rgb}{1,0.6,1}
\definecolor{rossochiaro}{rgb}{1,0.6,0.6}
\definecolor{verdechiaro}{rgb}{0.6,1,0.6}
\definecolor{giallochiaro}{rgb}{1,1,0.6}
\definecolor{bluscuro}{rgb}{0.15, 0.2, 0.9}
\definecolor{verdes}{rgb}{0.1, 0.5, 0.1}
\definecolor{light-gray}{gray}{0.7}

\makeatother   
\newcommand{\GeV}{{\rm \,GeV}}

 \def\be   {\begin{equation}}   \def\ee   {\end{equation}}
 \def\ba   {\begin{array}}      \def\ea   {\end{array}}
 \def\bea  {\begin{eqnarray}}   \def\eea  {\end{eqnarray}}
 \def\bean {\begin{eqnarray*}}  \def\eean {\end{eqnarray*}}

\renewcommand{\d}{\mathrm{d}}

\begin{document}

\vspace{7cm}

\begin{center}
\vspace{5cm}
{\Huge
Fermi Bubbles under Dark Matter Scrutiny\\ [0.9 cm]
Part II: Particle Physics Analysis
}
\\ [2.5cm]
{\large{\textsc{
Wei-Chih Huang$^{\,a,b,}$\footnote{\textsl{whuang@sissa.it}}, Alfredo Urbano$^{\,a,}$\footnote{\textsl{alfredo.urbano@sissa.it}}, Wei Xue$^{\,b,a,}$\footnote{\textsl{wxue@sissa.it}}}}}
\\[1cm]

\large{\textit{
$^{a}$~SISSA, via Bonomea 265, I-34136 Trieste, ITALY.\\ \vspace{1.5mm}
$^{b}$~INFN, sezione di Trieste, I-34136 Trieste, ITALY.
}}
\\ [2 cm]
{ \large{\textrm{
Abstract
}}}
\\ [1.5cm]
\end{center}

The analysis of the gamma-ray photons
collected by the Fermi Large Area Telescope reveals, after removal of astrophysical background,
the existence of an excess towards the Galactic center.
This excess peaks around few GeV, and its origin is compatible with the gamma-ray flux originating from Dark Matter annihilation.
In this work we take a closer look on this interpretation; we investigate which kind of Dark Matter, and which type of interactions
with the Standard Model fields are able to reproduce the observed signal. The structure of the paper is twofold.
In the first part, we follow an effective field theory approach considering both fermionic and scalar Dark Matter.
The computation of the relic density, the constraint imposed from the null result of direct searches, and the reliability
of the effective field theory description  allow us to single out only two viable dim-6 operators in the case of fermionic Dark Matter.
In the second part, we analyze some concrete models. In particular, we find that the scalar Higgs portal can provide a simple,
concrete and realistic scenario able to explain the GeV excess under scrutiny.

\def\thefootnote{\arabic{footnote}}
\setcounter{footnote}{0}
\pagestyle{empty}

\newpage
\pagestyle{plain}
\setcounter{page}{1}

\begin{addsections}


\subsection{Complex scalar Dark Matter}\label{sec:ScalarDM}

In this Section, we study the complex scalar DM which has the following structure
\be
   \mathcal{O}_\phi^i = \left\{ \bar{\phi} \phi  ,
      \ \ \partial_\mu \bar{\phi} \partial^\mu \phi ,
      \ \  \bar{\phi} \overset\leftrightarrow{\partial}_\mu \phi ,
      \ \  \partial_\mu \bar{\phi} \partial_\nu \phi\right\}~,
\ee
where $\bar{\phi} \overset\leftrightarrow{\partial}_\mu \phi \equiv \bar{\phi}(\partial_{\mu}\phi) - (\partial_{\mu}\bar{\phi})\phi$.

Similar to the fermionic DM, the effective operators in full generality can be written as
\bea
   {\rm Scalar:}~~~\mathcal{O}_{\rm S}^{s} &\equiv& \frac{m_f }{\sqrt{2}}~\bar{\phi}\phi~\bar{f}
      \left[ F_{\rm S}^{s}+F_{\rm SA}^{s}\gamma^5 \right]f~,
      \label{eq:SScalar}
   \\
   {\rm Vectorscalar:}~~~\mathcal{O}_{\rm VS}^{s}&\equiv&  \frac{m_f }{\sqrt{2}}~
      \partial_\mu \bar{\phi} \partial^\mu \phi~\bar{f} \left[
      F_{\rm VS}^{s}+F_{\rm VSA}^{s}\gamma^5 \right] f~,
      \label{eq:SVectorscalar}
\eea
\bea
   {\rm Vector:}~~~\mathcal{O}_{\rm V}^{s} &\equiv& \frac{i}{\sqrt{2}}~\bar{\phi} \overset\leftrightarrow{\partial}_{\mu}\phi
      ~\bar{f}\gamma^{\mu}\left[ F_{\rm V}^{s}+F_{\rm A}^{s}\gamma^5 \right]f~,
      \label{eq:SVector}
   \\
   {\rm Tensor:}~~~\mathcal{O}_{\rm T}^{s} &\equiv& \frac{m_f}{\sqrt{2}}~\partial^{[\mu} \bar{\phi}\partial^{\nu]}\phi
      ~\bar{f}\sigma_{\mu\nu}\left[ F_{\rm T}^{s}+F_{\rm TA}^{s}\gamma^{5} \right]f~,
      \label{eq:STensor}
\eea
where the antisymmetric combination $\partial^{[\mu} \bar{\phi}\partial^{\nu]}\phi \equiv
\partial^{\mu} \bar{\phi}\partial^{\nu}\phi - \partial^{\nu} \bar{\phi}\partial^{\mu}\phi$ preserves hermiticity.
Notice, moreover, that all the terms with $\gamma^5$ are CP-violating.

We analyze these operators following the same criteria adopted in the fermionic case.
\begin{table}[!htb!]
\begin{center}
\begin{tabular}[t]{|c||c|c|c|c|c|}
\hline
\multicolumn{6}{ |c| }{\textbf{Complex Scalar Dark Matter}}   \\ [1 pt] \hline \hline
\multirow{2}{*}{{{\color{bluscuro}{Operator}}}} & \multirow{2}{*}{{{\color{bluscuro}{Channel}}}} & \multicolumn{2}{ |c| }{{\color{bluscuro}{Annihilation cross section}}} &
\multirow{2}{*}{{{\color{bluscuro}{DD cross section}}}} &  \multirow{2}{*}{{{\color{bluscuro}{$s/\Lambda^2$ (\%)}}}}  \\ [1 pt] \cline{3-4}
& &  {{\color{bluscuro}{$m_f^2$ suppression}}}  &  {{\color{bluscuro}{$v^2$ suppression}}} &  &  \\ [1 pt]
\hline\hline   \rowcolor{light-gray}
 \cellcolor{white}  \multirow{4}{*}{{{\color{bluscuro}{S}}}} &  $\tau^+\tau^-$  (76.3) & \multirow{4}{*}{$\surd$}  & \multirow{4}{*}{$\times$} & $\times$  & 2.5 \\  \rowcolor{violachiaro}
 \cellcolor{white} & $c\bar{c}$    (58.2)    &  \multirow{4}{*}{}              & \multirow{4}{*}{}             & $\surd$  & 15.7  \\  \rowcolor{violachiaro}
 \cellcolor{white} & $b\bar{b}$   (57.5)     & \multirow{4}{*}{}               & \multirow{4}{*}{}             & $\surd$  & 34.8  \\   \rowcolor{rossochiaro}
 \cellcolor{white}  \multirow{-4}{*}{{{\color{bluscuro}{\textbf{S}}}}} & $q\bar{q}$  & \multirow{-4}{*}{$\surd$}  &  \multirow{-4}{*}{$\times$}  &  $\surd$  &  \\  [1 pt]
 \hline\hline   \rowcolor{light-gray}
 \cellcolor{white} \multirow{4}{*}{{{\color{bluscuro}{VS}}}}
 \cellcolor{white} & $\tau^+\tau^-$ (76.3) & \multirow{4}{*}{$\surd$}  & \multirow{4}{*}{$\times$} & $\times$  &  31.8  \\  \rowcolor{violachiaro}
 \cellcolor{white} & $c\bar{c}$  (58.2)       &  \multirow{4}{*}{}              & \multirow{4}{*}{}             & $\surd$  & 76  \\ \rowcolor{violachiaro}
 \cellcolor{white} & $b\bar{b}$ (57.5)       & \multirow{4}{*}{}               & \multirow{4}{*}{}             & $\surd$   & 118  \\  \rowcolor{rossochiaro}
 \cellcolor{white}  \multirow{-4}{*}{{{\color{bluscuro}{\textbf{VS}}}}} & $q\bar{q}$       & \multirow{-4}{*}{$\surd$}  &\multirow{-4}{*}{$\times$} & $\surd$  &   \\ [1 pt]
 \hline\hline  \rowcolor{rossochiaro}
 \cellcolor{white} \multirow{4}{*}{{{\color{bluscuro}{V}}}}
 \cellcolor{white} &  $\tau^+\tau^-$  & \multirow{4}{*}{$\times$}  & \multirow{4}{*}{$\surd$} & $\surd$ (1L)  &    \\  \rowcolor{rossochiaro}
 \cellcolor{white} & $c\bar{c}$      &  \multirow{4}{*}{}              & \multirow{4}{*}{}             & $\surd$ (1L) &    \\ \rowcolor{rossochiaro}
 \cellcolor{white} & $b\bar{b}$       & \multirow{4}{*}{}               & \multirow{4}{*}{}             & $\surd$ (1L)  &  \\ \rowcolor{rossochiaro}
 \cellcolor{white} \multirow{-4}{*}{{{\color{bluscuro}{\textbf{V}}}}} & $q\bar{q}$      &  \multirow{-4}{*}{$\times$} & \multirow{-4}{*}{$\surd$}& $\surd$ &    \\ [1 pt]
 \hline\hline \rowcolor{rossochiaro}
 \cellcolor{white} \multirow{4}{*}{{{\color{bluscuro}{T}}}}
 \cellcolor{white} & $\tau^+\tau^-$  & \multirow{4}{*}{$\surd$}  & \multirow{4}{*}{$\surd$} & \multirow{4}{*}{$\times$} &  \\   \rowcolor{rossochiaro}
 \cellcolor{white} & $c\bar{c}$        &  \multirow{4}{*}{}              & \multirow{4}{*}{}      & \multirow{4}{*}{}  &    \\   \rowcolor{rossochiaro}
 \cellcolor{white} & $b\bar{b}$       & \multirow{4}{*}{}               & \multirow{4}{*}{}             & \multirow{4}{*}{}  &    \\  \rowcolor{rossochiaro}
 \cellcolor{white} \multirow{-4}{*}{{{\color{bluscuro}{\textbf{T}}}}} & $q\bar{q}$    & \multirow{-4}{*}{$\surd$}             &\multirow{-4}{*}{$\surd$}&\multirow{-4}{*}{$\times$} &  \\ [1 pt] \hline
\end{tabular}
\end{center}
\caption{\textit{
The same as in Table.~\ref{tab:FermionDMproprieties}, but considering complex scalar DM according to the effective operators in Eqs.~(\ref{eq:SScalar}-\ref{eq:STensor}).
}}
\label{tab:ScalarDMproprieties}
\end{table}
With the same color code employed in Table~\ref{tab:FermionDMproprieties}, we summarize in Table~\ref{tab:ScalarDMproprieties} the main properties of the effective operators in Eqs.~(\ref{eq:SScalar}-\ref{eq:STensor}) associated with the DM annihilation cross section (relevant for the computation of the relic abundance and the fit of the Fermi bubbles excess), and the SI elastic cross section on nuclei relevant for DD. We collect the
explicit formulae in Appendix~\ref{app:A}. Our main results are presented in Fig.~\ref{fig:ScalarDarkMatter},
where we show the confidence regions obtained from the fit of the  Fermi bubbles excess together with the
contour reproducing the correct DM relic density and the bound placed by the XENON100 experiment. We
summarize the content of Table~\ref{tab:ScalarDMproprieties} and Fig.~\ref{fig:ScalarDarkMatter} as follows.

\begin{figure}[!htb!]
   \begin{minipage}{0.4\textwidth}
   \centering
   \includegraphics[scale=0.65]{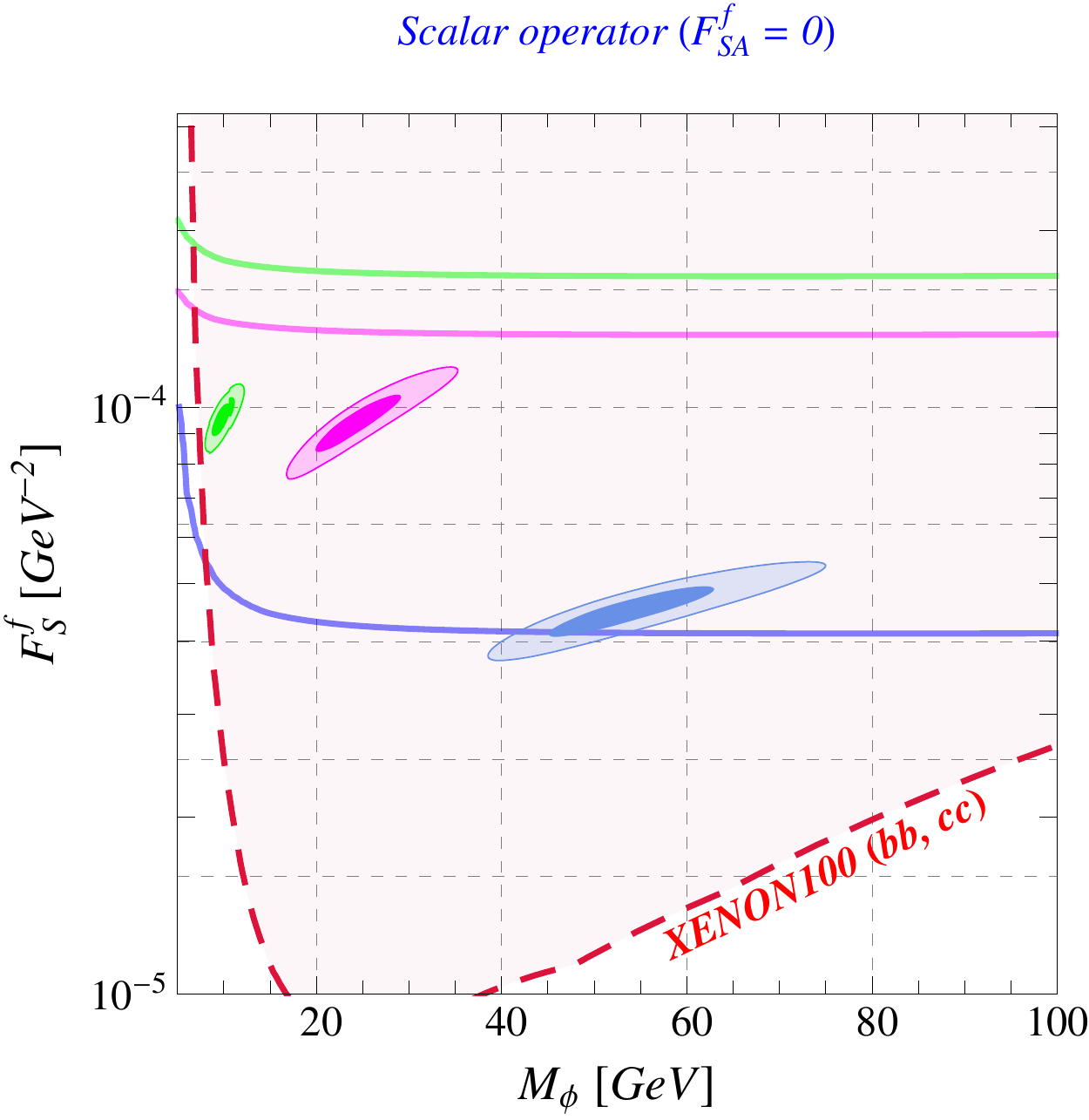}
    \end{minipage}\hspace{2 cm}
   \begin{minipage}{0.4\textwidth}
    \centering
    \includegraphics[scale=0.65]{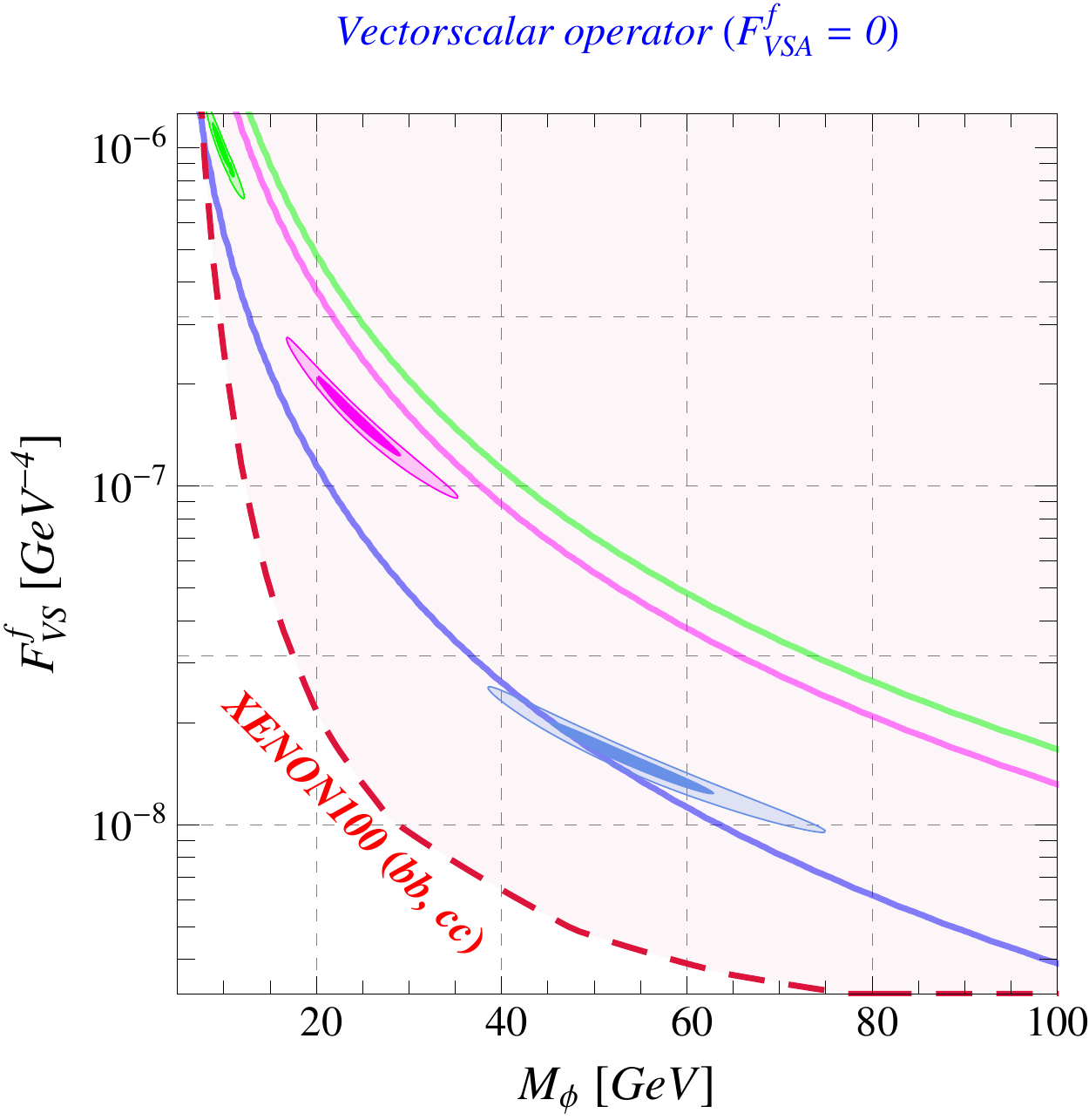}
    \end{minipage}
 \caption{\emph{The same as Fig.~\ref{fig:PseudoscalarPseudovectorOperators} but for complex scalar DM. In addition, we also show the bound obtained from the XENON100 experiment (red region excluded) considering $c\bar{c}$ and $b\bar{b}$ final states.}}
 \label{fig:ScalarDarkMatter}
 \end{figure}

\begin{itemize}
\item
{\underline{Scalar operator $\mathcal{O}_{\rm S}^{s}$}}. The annihilation cross section, both for the CP-preserving ($F_{\rm S}^{s}$) and CP-violating ($F_{\rm SA}^{s}$) operators, is mass suppressed but not velocity suppressed [see Eq.~(\ref{eq:ScalarXSectionS})], excluding the light quark channels. Let us consider for definiteness the case $F_{\rm SA}^{s}=0$. Final
states involving $\tau^+\tau^-$, $c\bar{c}$, $b\bar{b}$
provide a good fit to the Fermi bubbles excess, as shown in the left panel of Fig.~\ref{fig:ScalarDarkMatter}, where we plot the 68~\% and 99~\% confidence regions. The $b\bar{b}$ final state is the only channel
having a overlap between the confidence region and the contour reproducing the observed value of relic density. Notice, in addition, that the effective field theory approach
with   $c\bar{c}$ and $b\bar{b}$ final states
fails to provide a good and satisfactory description, introducing an error always larger than our nominal 10~\% threshold of acceptance.
Finally,
in terms of DD searches,
the scalar operator with $c\bar{c}$ and $b\bar{b}$ final states has a large SI elastic cross section.
This is because, via a triangle loop, DM can interact with the gluon content of the nucleus. We briefly review
the computation of the resulting cross section in Appendix~\ref{App:TreeLevelDD}. It turns out that the XENON100 experiment places a strong bound on the parameter space of the scalar operator. In particular, the region favored by
 the fit of the Fermi bubbles excess and the contour reproducing the correct relic density are ruled out.

Let us consider the opposite situation, i.e. with $F_{\rm S}^{s}=0$,  $F_{\rm SA}^{s}\neq 0$; the constraint from the XENON100 experiment goes away because the SI cross section for pseudoscalar interactions is zero. As a consequence, the annihilation into $b\bar{b}$ could provide a good fit to the Fermi bubbles excess and yield the correct relic density.\footnote{The reason why we do not show the analogous of Fig.~\ref{fig:ScalarDarkMatter} for this case $F_{\rm S}^{s}=0$, $F_{\rm SA}^{s}\neq 0$ is that, as shown in Section~\ref{sec:FermionicDM}, the two cases differ only by subleading corrections $\mathcal{O}(m_f^2/M_{\phi}^2)$, thus leaving unchanged the results of the confidence region and the DM contour.}
Notice that, however, the effective field theory approach fails to provide a good approximation due to large $s/\Lambda^2$.

To sum, we argue that an effective field theory description of complex scalar DM annihilation
based on the scalar operator in Eq.~(\ref{eq:SScalar}) fails to explain the Fermi bubbles excess.

In Section~\ref{sec:MSSM_Higgs}, we shall see a concrete realization beyond the effective field theory approach of
Eq.~(\ref{eq:SScalar}), in which all the tensions appearing in this analysis will be overcome.

\item
{\underline{Vectorscalar operator $\mathcal{O}_{\rm VS}^{s}$}}. The phenomenological
analysis mimics the scalar case. As far as concerns the computation of the annihilation cross section, in fact,
the derivatives on the DM side of the effective operator
 pick up a factor $k_1 \cdot k_2 \approx M_{\phi}^2$ at the amplitude level, where $k_{1,2}$ are the incoming DM particle four-momenta [see Eq.~(\ref{eq:ScalarXSectionVS}) for the corresponding cross section].
Both the confidence region obtained from the fit of the Fermi bubbles excess and the contour
reproducing the correct DM relic density are just shifted by this factor, as shown in the right panel of Fig.~\ref{fig:ScalarDarkMatter}. We observe the same behavior
also in the computation of the elastic cross section on nuclei; in this case, in fact, the kinematic of the elastic scattering
implies $2 k_1 \cdot k_2 = -q^2 + 2M_{\phi}^2\approx 2M_{\phi}^2$, where $q$ is the transferred momentum.

As a consequence, we reach the same conclusion as in the scalar operator: the vectorscalar operator
fails to explain the Fermi bubbles excess.

\item

{\underline{Vector operator $\mathcal{O}_{\rm V}^{s}$ and Tensor operator $\mathcal{O}_{\rm T}^{s}$}}. The annihilation cross sections are mass and velocity suppressed, as shown explicitly in Eqs.~(\ref{eq:ScalarXSectionV},\ref{eq:ScalarXSectionT}). These operators can not reproduce the Fermi bubbles excess due to the low DM velocity ($v \sim 10^{-3}$).

\end{itemize}

\end{addsections}

\begin{toexclude}

\section{Introduction}

The quest for the first non-gravitational evidence  of Dark Matter (DM) in the Galaxy is becoming, as time goes by, one of the most intriguing and challenging tasks in astrophysics.

Paradoxically though it may seem -- DM should be dark, after all -- one of the most promising signals that could reveal the
 elusive fingerprints
of DM  is the analysis of the gamma-ray photons collected by the Fermi Large Area Telescope (LAT).
A number of recent analysis, in particular,
point towards the possible existence of a residual excess originating from the Galactic Center (GC), peaked at few GeV, and compatible with the annihilation of DM particles with a thermally averaged cross section, $\langle \sigma v\rangle \sim 10^{-26}$ cm$^3$s$^{-1}$~\cite{Hooper:2011ti,Hooper:2010mq,Dodelson:2007gd,Boyarsky:2012ca,Abazajian:2012pn}, that is of the same order as the one required if DM is in thermal equilibrium in the early  Universe.

Remarkably, a very similar excess has been found in Refs.~\cite{Hooper:2013rwa,Huang:2013pda} studying the gamma-ray spectrum originating from  the so-called Fermi bubbles. The name Fermi bubbles refers to a colossal pair
of lobe-shaped structures, first observed in Refs.~\cite{Dobler:2009xz, Su:2010qj}, each extending tens of thousands of light-years above and below the Galactic plane, and covering more than one half of the visible sky, from the constellation of Virgo to the constellation of Grus. Invisible to the naked eye, they reveal themselves in full glory in the energy region from 300 MeV up to 300 GeV.
Focusing the analysis of the corresponding energy spectrum on high latitudes -- i.e. for $|b|>30^{\circ}$, where $b$ is the latitude in Galactic coordinates with $b=0^\circ$
corresponding to the Galactic plane -- the Fermi bubbles present a fairly flat (in $E_{\gamma}^2 d\Phi/dE_{\gamma}d\Omega$) spectrum, with a magnitude around $3\times 10^{-7}$ GeV/cm$^2$/s/sr. A common explanation relies on the assumption of the existence of a spatially extended population of high-energy electrons trapped inside the bubbles, emitting gamma rays via Inverse Compton Scattering (ICS) on the ambient light. Moreover, this hypothesis allows to establish, via synchrotron radiation in the presence of microgauss magnetic field, an interesting spatial correlation with the WMAP haze~\cite{Dobler:2009xz} observed in the microwaves.

The key observation made in Refs.~\cite{Hooper:2013rwa,Huang:2013pda}, however, is that
at low latitudes, in particular for $|b|= 10^{\circ}-20^{\circ}$,  the energy spectrum
of the Fermi bubbles peaks at $E_{\gamma}\sim 1 - 4$ GeV, thus revealing the possible existence of an extra component that, in addition to the aforementioned ICS photons, becomes distinguishable toward the GC. Refs.~\cite{Hooper:2013rwa,Huang:2013pda} have shown that the extra component
 can be explained by DM annihilation, and the best fit obtained in Ref.~\cite{Huang:2013pda} corresponds to annihilation into $b\bar{b}$ with mass $M_{\rm DM}=61.6$ GeV and thermally averaged annihilation cross section $\langle \sigma v\rangle = 3.38\times 10^{-26}$
 cm$^3$s$^{-1}$.  Besides, Ref.~\cite{Hooper:2013nhl} excludes the millisecond Pulsar explanation.

  Assuming the validity of the DM interpretation,
 in this paper we analyze the excess
 found in Refs.~\cite{Hooper:2013rwa,Huang:2013pda} from
 the particle physics perspective.

Starting from the assumption that DM is a Weakly Interacting Massive Particle (WIMP), the best-suited tool to perform this analysis is the framework provided by the effective field theories.
This approach, integrating out the details of DM interactions at small distances, has the benefit of capturing a model-independent picture of the scrutinized signal, reaching general conclusions that can be used as guidelines for more complicated and concrete models. Pursuing this goal, we investigate both fermionic and complex scalar DM. First,
assuming that DM is a singlet under the Standard Model (SM) gauge group,
we construct in full generality the effective operators describing
at leading order the interactions between DM and the SM fields.
Second, we perform a careful phenomenological analysis taking into account, in addition to the fit of the aforementioned Fermi bubbles signal, the requirement to reproduce the correct relic density as recently measured by the Planck collaboration \cite{Ade:2013zuv}, and the bound imposed by the null result of the XENON100 experiment in the context of Direct Detection (DD) of DM \cite{Aprile:2012nq}. For the analysis of the DD constraint we include
the elastic cross section generated at one-loop via electromagnetic interactions. In correspondence of each analyzed operators, moreover, we also comment about the reliability
of the effective field theory description.

From a complementary point of view, we complete our analysis presenting
 two concrete models that rely on the exchange of a light resonance between the dark sector and the SM fields, thus representing a setup that falls beyond the domain of the effective field theory description. In particular, we investigate
 the scalar Higgs portal and a generic model with an extra $Z^{\prime}$.

This paper is organized as follows. In Section \ref{sec:FermiBubbles}, we describe the data that are used throughout this work,
and related to the excess found in Ref.~\cite{Hooper:2013rwa,Huang:2013pda};
in order to further motivate the DM explanation, moreover, we perform a model-independent fit using as free parameters the DM mass and the thermally averaged annihilation cross section.
In Section~\ref{sec:EFTapproach}, we go one step further in the particle physics analysis, presenting
 and discussing the approach based on the effective operators. Section~\ref{sec:MSSM} is devoted to the concrete realizations. Finally we conclude
in Section \ref{sec:Conclusions}. In Appendix~\ref{app:0}, we present the gamma-ray data used throughout the paper.
We collect all the relevant analytical formulas in Appendix~\ref{app:A}, while
we quickly review the integration of the Boltzmann equation in Appendix~\ref{App:Boltzmann}.

\section{On the presence of a Dark Matter component in the Fermi bubbles energy spectrum}\label{sec:FermiBubbles}

In this Section, we describe in detail the data used in this paper. As already mentioned in the Introduction,
the analysis performed in Refs.~\cite{Hooper:2013rwa,Huang:2013pda}
reveals that the energy spectrum of the Fermi bubbles arises from the combination of two different components:
\textit{i}) an ICS component, dominant at high latitudes, produced by a population of high-energy electrons trapped inside the bubbles, and \textit{ii}) an additional component, responsible for a bump at $E_{\gamma}\sim 1-4$ GeV, compatible with DM annihilation, and dominant at low latitudes. In Ref.~\cite{Huang:2013pda} these two component have been studied together, performing a fit of the whole Fermi bubbles energy spectrum in the energy range $E_{\gamma}=0.3-300$ GeV. In this paper, however, we are mostly interested in the analysis of the DM component. In order to simplify the analysis, as a consequence, we subtract the ICS component from the energy spectrum of the Fermi bubbles; this subtraction procedure, in fact, allows us to isolate the DM contribution in which we are interested.
We follow the approach provided by Ref.~\cite{Hooper:2013rwa}, and we present in Appendix~\ref{app:0} the resulting data (collected in Table~\ref{tableDATA1} and Table~\ref{tableDATA2}).
 As a representative example, we show in the left panel of Fig.~\ref{fig:BubbleFit}
 the energy spectrum of the Fermi bubbles after ICS subtraction in the region $|b|=10^{\circ}-20^{\circ}$ where
 the bump at $E_{\gamma}\sim 1-4$ GeV clearly stands out. In the rest of the paper, we will refer to the spectrum of the Fermi bubbles after ICS subtraction as the ``Fermi bubbles excess''.

We are now in the position to fit the data describing the Fermi bubbles excess using the prompt gamma rays produced from DM annihilation.\footnote{As noticed in Ref.~\cite{Hooper:2013rwa,Huang:2013pda}, the ICS photons produced by DM annihilation do not play an important role in the interpretation of the Fermi bubbles excess.} Our purpose is to capture, in a completely model-independent way, the most general
features of this signal in order to have a guideline for the rest of the analysis. The differential photon flux from DM annihilation is given by
\begin{equation}
  \frac{d\Phi_{\gamma}}{dE_{\gamma} d\Omega}=\frac{r_{\odot}}{4\pi}
  \frac{1}{2c}\left(\frac{\rho_{\odot}}{M_{\rm DM}}\right)^2
   \bar{J}  \sum_f \langle\sigma v \rangle_f \frac{dN_{\gamma}^{f}}{dE_{\gamma}} \ ,
\label{eq:Flux}
\end{equation}
where $dN_{\gamma}^{f}/dE_{\gamma}$
 is the number of photon per unit energy per DM
annihilation with final state $f$ and thermally averaged cross section $\langle\sigma v \rangle_f$, with $c=1$ ($c=2$) for Majorana (Dirac) DM;
$\bar{J}$ is the angular average of the $J$ factor, $\bar{J}=(1/\Delta\Omega)\int\int db dl \cos b\, J(\theta)$,
where
\begin{equation}
J(\theta)=\int_{\rm l.o.s.}\frac{ds}{r_{\odot}}\left[\frac{\rho_{\rm gNFW}(r(s,\theta))}{\rho_{\odot}}\right]^{2}~,
\end{equation}
and $r(s,\theta)=(r_{\odot}^2+s^2-2r_{\odot}s\cos\theta)^{1/2}$.
Following Refs.~\cite{Hooper:2013rwa,Huang:2013pda}, we use the generalized NFW profile with an inner slope $\gamma  = 1.2$
\footnote{The standard NFW profile has an inner slope of $\gamma=1$.}
\begin{equation}\label{eq:NFW}
\rho_{\rm gNFW}(r)=\rho_s\left(\frac{r}{R_s}\right)^{-\gamma}
\left(1+\frac{r}{R_s}\right)^{\gamma-3}~,
\end{equation}
where the scale radius is $R_s = 20$ kpc, and at the location of the Sun
$r_{\odot}=8.33$ kpc the DM density $\rho_{\odot}$ is normalized to $0.4$ GeV/cm$^3$. The values of the averaged $J$ factor used in our analysis are listed in Appendix~\ref{app:0}.
\begin{figure}[!htb!]
  \begin{minipage}{0.4\textwidth}
   \centering
   \includegraphics[scale=0.65]{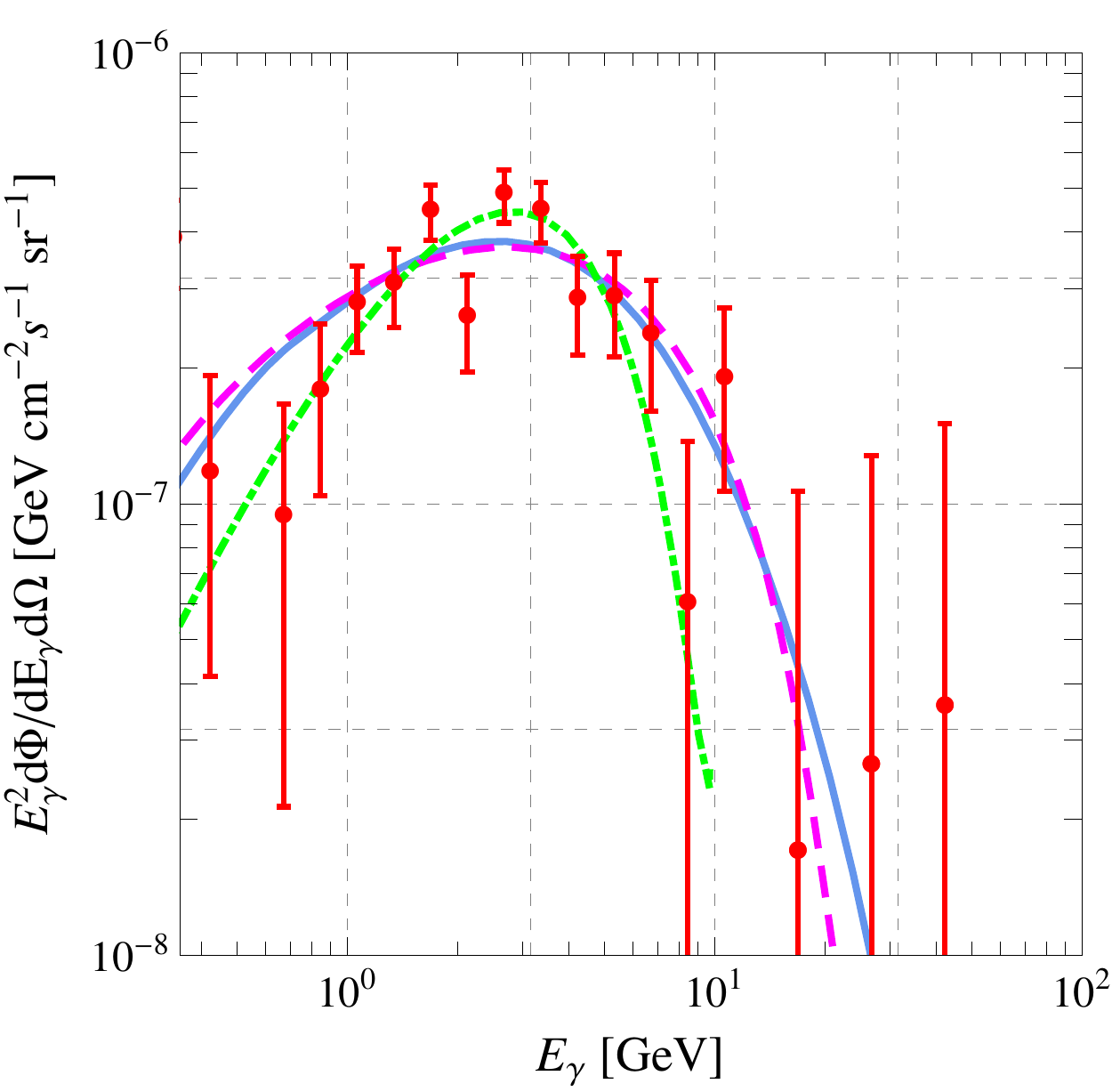}
    \end{minipage}\hspace{1.65 cm}
  \begin{minipage}{0.4\textwidth}
   \centering
   \includegraphics[scale=0.65]{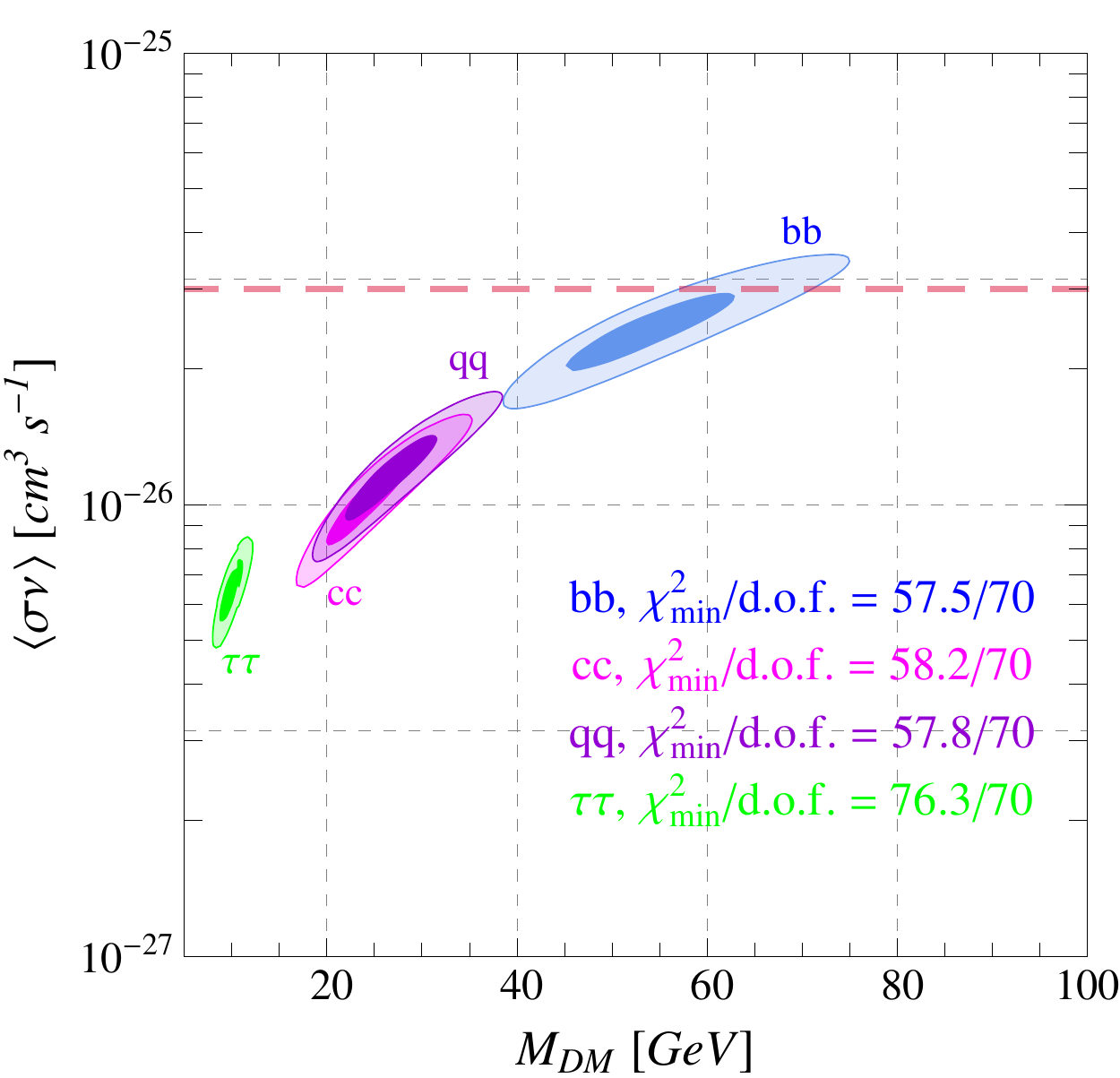}
    \end{minipage}\hspace{2 cm}\\
 \caption{\emph{Left panel: energy spectrum of the Fermi bubbles after ICS subtraction in the region $|b|=10^{\circ}-20^{\circ}$. We also show the photon spectrum produced by DM
 annihilation into $b\bar{b}$ (solid blue line), $c\bar{c}$ (dashed magenta line), $\tau^+\tau^-$ (dot-dashed green line) in correspondence of the best-fit values for $M_{\rm DM}$ and $\langle \sigma v\rangle$. Right panel: confidence regions (68~\% C.L. and 95~\% C.L.) in the plane $[M_{\rm DM},\langle \sigma v\rangle]$ obtained from the chi-square
 fit of the Fermi bubbles excess using the prompt gamma-ray flux produced by DM annihilation in Eq.~\ref{eq:Flux}. We explore
 different annihilation channels, i.e. $b\bar{b}$, $c\bar{c}$, $q\bar{q}$, $\tau^+\tau^-$. The goodness of the fit can be characterized by means of the reduced chi-square $\chi_{\rm min}^{2}/ n$, where $n \equiv N-p$, $N$ is the number of data points of the non-negative photon flux, and $p=2$ is the number of the fitting parameters. The dashed red line corresponds to the value $\langle \sigma v\rangle = 3\times 10^{-26}$ cm$^3$s$^{-1}$.}}
 \label{fig:BubbleFit}
\end{figure}
We perform a chi-square analysis, and we show our results in the right panel of Fig.~\ref{fig:BubbleFit}.
For simplicity, we study all the possible two-body final states one-by-one,
 i.e. assuming 100~\% DM annihilation into each one of the SM channels.
We find that
the only annihilation channels that can fit the Fermi bubbles excess involve $b\bar{b}$, $c\bar{c}$, $q\bar{q}$, $\tau^+\tau^-$ in the final state. The photon spectrum produced by DM annihilation into
 final states involving light leptons, electroweak gauge bosons and the Higgs, in particular,
 can not reproduce the bump observed at $E_{\gamma}\sim 1-4$ GeV. The best fit values for the DM mass and the thermally averaged
cross section vary from
 $M_{\rm DM}\sim 10$ GeV, $\langle \sigma v\rangle\sim 6\times 10^{-27}$ cm$^3$s$^{-1}$ (annihilation into $\tau^+\tau^-$) to
 $M_{\rm DM}\sim 60$,  $\langle \sigma v\rangle\sim 2\times 10^{-26}$ cm$^3$s$^{-1}$ GeV (annihilation into $b\bar{b}$).

Let us now discuss the implication of Fig.~\ref{fig:BubbleFit}
from the point of view of the DM relic density.
The evolution of the DM density, according to the standard freeze-out scenario, is driven by the
expansion of the Universe and the
interactions of DM with SM particles;  this picture finds a quantitative description in terms of a Boltzmann equation whose approximate solution is
\begin{equation}\label{eq:MasterRelic}
\frac{\Omega_{\rm DM}h^2}{0.1199} \approx \frac{3\times 10^{-26}~{\rm cm}^3{\rm s}^{-1}}
{\langle \sigma v\rangle}~.
\end{equation}
The value of the DM relic abundance measured by the Planck collaboration \cite{Ade:2013zuv} is $\Omega_{\rm DM}h^2 = 0.1199\pm
0.0027$. As a consequence, from the right panel of Fig.~\ref{fig:BubbleFit},
one can naively conclude that only the $b\bar{b}$ final state has the possibility to
reproduce the  correct value of relic density within the confidence region obtained from the fit of the Fermi bubbles excess.
This intuition is true, however, only if the thermally averaged cross section in Eq.~(\ref{eq:Flux}) and Eq.~(\ref{eq:MasterRelic}) are the same.
Considering the expansion, $\langle \sigma v\rangle= a + bv^2 +\mathcal{O}(v^4)$,
 in Eq.~(\ref{eq:Flux}) we have $\langle \sigma v\rangle\approx a$, since at the present
  $v^2 \sim 10^{-6}$.
In order to have the same thermally averaged cross section
in Eq.~(\ref{eq:MasterRelic}),
therefore,
the s-wave must dominate over the p-wave also during the freeze-out epoch,
i.e. $a \gg  bv^2$, with $v^2 \sim 1/4$.
This request is far from obvious,
and depends on the details of the interaction between DM and SM fermions. To fully understand
the connection between the relic density and the fit of the Fermi bubbles excess, therefore,
we are forced to abandon the model-independent perspective pursued in this Section.
In the next Section, we shall investigate the interactions between DM and SM fermions using an effective field theory approach; we shall encounter a situation in which the p-wave is not negligible
at the freeze-out epoch, thus drastically modifying the scenario depicted in the right panel of Fig.~\ref{fig:BubbleFit}.

 Before proceeding to our analysis, a caveat is mandatory.
A different DM density profile other than the gNFW, used
throughout this paper, results in different values for the averaged $J$-factor in Eq.~(\ref{eq:Flux}).
For instance, going from the gNFW to the NFW profile, the averaged $J$-factor
turns out to be reduced by $30~\%$.
As a consequence, in the fit of the Fermi bubbles excess, the
favored value of the thermally averaged cross section increases in
order to counterbalance this effect.

\section{Effective field theory approach}\label{sec:EFTapproach}
In the previous Section, we have studied the Fermi bubbles excess
treating the thermally averaged annihilation cross section $\langle \sigma v \rangle$
as a free parameter, i.e.
without detailing the interactions between DM and the SM sector. Taking one step further,
in this Section
we explore to what extent the possibility to reproduce the signal depends on the structure
of these interactions with the help of the effective
field theory approach.

Pursuing this direction, we start with the following assumptions:
\begin{enumerate}
\item  DM is a WIMP, and it is non-relativistic when it decouples from thermal equilibrium, i.e. it is cold DM;

\item DM is a singlet under the SM gauge group;

\item any new particle beyond the SM is much heavier than the DM particle; this assumption excludes resonance enhancement and/or co-annihilation processes.

\end{enumerate}

In the following, we consider both fermionic (Section~\ref{sec:FermionicDM}) and complex scalar (Section~\ref{sec:ScalarDM}) DM.

\subsection{Fermionic Dark Matter}\label{sec:FermionicDM}

In the framework delimited by our assumptions, the effective operators describing the interactions between DM and the SM particles take the following factorized form
\begin{equation}
\mathcal{O}_{\chi}^{i}\times \mathcal{O}_{\rm SM}^{i}~,
\end{equation}
where $i$ explicitly implies the contraction over Lorentz indices. Driven by the results obtained in Section~\ref{sec:FermiBubbles}, we neglect heavy final states involving $W^{\pm}$, $Z$, and the Higgs boson. In Section~\ref{sec:MSSM}, we will loosen some of the assumptions for specific benchmark models.

We study the following DM structures
\begin{equation}
\mathcal{O}_{\chi}^{i}=\bar{\chi}\Gamma^{i}\chi~,~~~~~\Gamma^{i}=\left\{\textbf{1},\gamma^{5}, \gamma^{\mu}, \gamma^{\mu}\gamma^{5}, \sigma^{\mu\nu}\right\}~,
\end{equation}
where DM  $\chi$ can be either Majorana or Dirac except for the vector and tensor case, in which only Dirac DM is present.
From the SM side, we couple $\mathcal{O}_{\chi}^{i}$ to gauge invariant SM currents. Considering for definiteness the vector-like interaction $\bar{\chi}\Gamma^{\mu}\chi$, this means that we take
\begin{eqnarray}
\mathcal{O}_{\rm SM, V}^{\mu}&=& \frac{G_{\rm V,L}^{f}}{\sqrt{2}}(\bar{f}_{1,L}~\bar{f}_{2,L})\gamma^{\mu}
\left(
\begin{array}{c}
  f_{1,L}  \\
  f_{2,L}
\end{array}
\right)+\frac{G_{\rm V,R}^{f_1}}{\sqrt{2}}\bar{f}_{1,R}\gamma^{\mu}f_{1,R}
+\frac{G_{\rm V,R}^{f_2}}{\sqrt{2}}\bar{f}_{2,R}\gamma^{\mu}f_{2,R}\nonumber\\
&=&\frac{1}{\sqrt{2}}\sum_{i=1,2}\bar{f}_i\gamma^{\mu}\left[
G_{\rm V}^{f_i}+G_{\rm A}^{f_i}\gamma^5\right]f_i~,\label{eq:SMVector}
\end{eqnarray}
where $G_{\rm V}^{f_i}\equiv (G_{\rm V,L}^{f}+G_{\rm V,R}^{f_i})/2$, $G_{\rm A}^{f_i}\equiv (-G_{\rm V,L}^{f}+G_{\rm V,R}^{f_i})/2$. In Eq.~(\ref{eq:SMVector}), $(f_{1,L}~f_{2,L})^T$ is the SM $SU(2)_L$ doublet, while $f_{i,R}$ are the corresponding singlets. All in all we find in full generality\footnote{We consider only effective operators at dim-6. See Ref.~\cite{DeSimone:2013gj} for a recent discussion about the relevance of dim-8 operators for a TeV-scale DM particle.}
\begin{eqnarray}
{\rm Scalar:}~~~\mathcal{O}_{\rm S}^{f} &\equiv& \frac{m_f }{\sqrt{2}}~\bar{\chi}\chi~\bar{f}
\left[
G_{\rm S}^{f}+G_{\rm SA}^{f}\gamma^5
\right]f~,\label{eq:Scalar}
\\
{\rm Pseudoscalar:}~~~\mathcal{O}_{\rm PS}^{f}&\equiv&  \frac{m_f }{\sqrt{2}}~
\bar{\chi}\gamma^5\chi~\bar{f}
\left[
G_{\rm PS}^{f}+G_{\rm PSA}^{f}\gamma^5
\right]
f~,\label{eq:Pseudoscalar}\\
{\rm Vector:}~~~\mathcal{O}_{\rm V}^{f} &\equiv& \frac{1}{\sqrt{2}}~\bar{\chi}\gamma^{\mu}\chi
~\bar{f}\gamma_{\mu}\left[
G_{\rm V}^{f}+G_{\rm VA}^{f}\gamma^5
\right]f~,\label{eq:Vector}\\
{\rm Pseudovector:}~~~\mathcal{O}_{\rm PV}^{f} &\equiv& \frac{1}{\sqrt{2}}~\bar{\chi}\gamma^{\mu}\gamma^{5}\chi
~\bar{f}\gamma_{\mu}\left[
G_{\rm PV}^{f}+G_{\rm PVA}^{f}\gamma^5
\right]f~,\label{eq:Pseudovector}\\
{\rm Tensor:}~~~\mathcal{O}_{\rm T}^{f} &\equiv& \frac{m_f}{\sqrt{2}}~\bar{\chi}\sigma^{\mu\nu}\chi~\bar{f}\sigma_{\mu\nu}\left[
G_{\rm T}^{f}+G_{\rm TA}^{f}\gamma^{5}
\right]f~,\label{eq:Tensor}
\end{eqnarray}
where $\sigma^{\mu\nu}\equiv i[\gamma^{\mu},\gamma^{\nu}]/2$.
Notice that operators with
$G_{\rm SA}^{f}$ in Eq.~(\ref{eq:Scalar}),
$G_{\rm PS}^{f}$ in Eq.~(\ref{eq:Pseudoscalar}) and
$G_{\rm TA}^{f}$ in Eq.~(\ref{eq:Tensor})
  are CP-violating.\footnote{In Eq.~(\ref{eq:Tensor}), we have the following CP transformation property
  \begin{equation}
  \mathcal{O}_{\mu\nu}^{5}\equiv \bar{f}\frac{i}{2}(\gamma_{\mu}\gamma_{\nu}-
  \gamma_{\nu}\gamma_{\mu})\gamma^5f  \overset{\rm CP}{\Longrightarrow} (-1)^{\mu}(-1)^{\nu}
  \mathcal{O}_{\mu\nu}^{5},
  \end{equation}
  where $(-1)^0=1$ and $(-1)^i=-1$ for $i=1,2,3$.}
 In the absence of CP violation in the DM sector, these operators are zero. We, nevertheless, include them in our analysis because of lack of knowledge in the DM sector.
 The mass insertion in Eqs.~(\ref{eq:Scalar}, \ref{eq:Pseudoscalar}, \ref{eq:Tensor}) manifests the involvement of the Higgs doublet to break the chiral symmetry without violating the gauge symmetry, like the SM Yukawa couplings.

\begin{table}[!htb!]
\begin{center}
\begin{tabular}[t]{|c||c|c|c|c|c|}
\hline
\multicolumn{6}{ |c| }{\textbf{Fermionic Dark Matter}}   \\ [1 pt] \hline \hline
\multirow{2}{*}{{{\color{bluscuro}{Operator}}}} & \multirow{2}{*}{{{\color{bluscuro}{Channel}}}} & \multicolumn{2}{ |c| }{{\color{bluscuro}{Annihilation cross section}}} &
\multirow{2}{*}{{{\color{bluscuro}{DD cross section}}}} &  \multirow{2}{*}{{{\color{bluscuro}{$s/\Lambda^2$ (\%)}}}}  \\ [1 pt] \cline{3-4}
& &  {{\color{bluscuro}{$m_f^2$ suppression}}}  &  {{\color{bluscuro}{$v^2$ suppression}}} &  &   \\ [1 pt] \hline\hline   \rowcolor{rossochiaro}
 \cellcolor{white}  \multirow{4}{*}{{{\color{bluscuro}{S}}}} &  $\tau^+\tau^-$ & \multirow{4}{*}{$\surd$}  & \multirow{4}{*}{$\surd$} & $\times$  & \\  \rowcolor{rossochiaro}
 \cellcolor{white} & $c\bar{c}$        &  \multirow{4}{*}{}              & \multirow{4}{*}{}             & $\surd$  &  \\  \rowcolor{rossochiaro}
 \cellcolor{white} & $b\bar{b}$       & \multirow{4}{*}{}               & \multirow{4}{*}{}             & $\surd$  &   \\  \rowcolor{rossochiaro}
 \cellcolor{white}  \multirow{-4}{*}{{{\color{bluscuro}{\textbf{S}}}}} & $q\bar{q}$  & \multirow{-4}{*}{$\surd$}  &  \multirow{-4}{*}{$\surd$}  &  $\surd$  &   \\  [1 pt] \hline\hline   \rowcolor{light-gray}
 \cellcolor{white} \multirow{4}{*}{{{\color{bluscuro}{PS}}}}
 \cellcolor{white} & $\tau^+\tau^-$ (76.3) & \multirow{4}{*}{$\surd$}  & \multirow{4}{*}{$\times$} &  \multirow{4}{*}{$\times$}  &  13.7  \\  \rowcolor{light-gray}
 \cellcolor{white} & $c\bar{c}$  (58.2)       &  \multirow{4}{*}{}              & \multirow{4}{*}{}             & \multirow{4}{*}{}  &  43.7  \\ \rowcolor{giallochiaro}
 \cellcolor{white} & $b\bar{b}$ (57.5)       & \multirow{4}{*}{}               & \multirow{4}{*}{}             & \multirow{4}{*}{}  &  78.5  \\ \rowcolor{rossochiaro}
 \cellcolor{white}  \multirow{-4}{*}{{{\color{bluscuro}{\textbf{PS}}}}} & $q\bar{q}$       & \multirow{-4}{*}{$\surd$}  &\multirow{-4}{*}{$\times$} & \multirow{-4}{*}{$\times$}  &   \\ [1 pt] \hline\hline  \rowcolor{light-gray}
 \cellcolor{white} \multirow{4}{*}{{{\color{bluscuro}{V}}}}
 \cellcolor{white} &  $\tau^+\tau^-$ (76.3)  & \multirow{4}{*}{$\times$}  & \multirow{4}{*}{$\times$} & $\surd$ (1L)  &  0.3  \\   \rowcolor{violachiaro}
 \cellcolor{white} & $c\bar{c}$ (58.2)        &  \multirow{4}{*}{}              & \multirow{4}{*}{}             & $\surd$ (1L) & 0.6  \\
 \cellcolor{white} & $b\bar{b}$ (57.5)       & \multirow{4}{*}{}               & \multirow{4}{*}{}             & $\surd$ (1L)  &  1.9 \\ \rowcolor{violachiaro}
 \cellcolor{white} \multirow{-4}{*}{{{\color{bluscuro}{\textbf{V}}}}} & $q\bar{q}$ (57.8)      &  \multirow{-4}{*}{$\times$} & \multirow{-4}{*}{$\times$}& $\surd$ &  0.7  \\ [1 pt] \hline\hline
 \cellcolor{white} \multirow{4}{*}{{{\color{bluscuro}{PV}}}}
 \cellcolor{white} & $\tau^+\tau^-$ (76.3) & \multirow{4}{*}{$\surd$}  & \multirow{4}{*}{$\times$} & \multirow{4}{*}{$\times$}  &  2.5 \\   \rowcolor{light-gray}
 \cellcolor{white} & $c\bar{c}$  (58.2)      &  \multirow{4}{*}{}              & \multirow{4}{*}{}             & \multirow{4}{*}{}  & 14.4  \\
\rowcolor{giallochiaro}
 \cellcolor{white} & $b\bar{b}$ (57.5)      & \multirow{4}{*}{}               & \multirow{4}{*}{}             & \multirow{4}{*}{}  & 34.6  \\
\rowcolor{rossochiaro}
 \cellcolor{white} \multirow{-4}{*}{{{\color{bluscuro}{\textbf{PV}}}}} & $q\bar{q}$    & \multirow{-4}{*}{$\surd$}             &\multirow{-4}{*}{$\times$}&\multirow{-4}{*}{$\times$} &   \\ [1 pt] \hline\hline  \rowcolor{light-gray}
 \multirow{4}{*}{{{\color{bluscuro}{T}}}}
 \cellcolor{white} & $\tau^+\tau^-$ (76.3) & \multirow{4}{*}{$\surd$}  & \multirow{4}{*}{$\times$} & \multirow{4}{*}{$\times$}  &  8.3  \\ \rowcolor{light-gray}
 \cellcolor{white} & $c\bar{c}$ (58.2)        &  \multirow{4}{*}{}              & \multirow{4}{*}{}             & \multirow{4}{*}{}  & 29.1  \\ \rowcolor{giallochiaro}
 \cellcolor{white} & $b\bar{b}$ (57.5)      & \multirow{4}{*}{}               & \multirow{4}{*}{}             & \multirow{4}{*}{}  &  49.1  \\ \rowcolor{rossochiaro}
 \cellcolor{white}  \multirow{-4}{*}{{{\color{bluscuro}{\textbf{T}}}}} & $q\bar{q}$      &  \multirow{-4}{*}{$\surd$}  & \multirow{-4}{*}{$\times$}& \multirow{-4}{*}{$\times$}&     \\ [1 pt] \hline
\end{tabular}
\end{center}
\caption{\textit{Properties of scattering cross sections considering  fermionic DM. For each operator we highlight the corresponding behavior concerning annihilation processes (3$^{\,\rm td}$ and 4$^{\rm th}$ column), and SI elastic cross sections on nuclei  (5$^{\rm th}$ and 6$^{\rm th}$ column).
The symbol $\surd$ ($\times$) marks a property possessed (not possessed) by the corresponding operator. The symbol {\rm 1L} indicates that the corresponding elastic cross section arises at one-loop level. We refer to the Appendix \ref{app:A} for the analytical expressions. We focus our analysis on final states of $\tau^+\tau^-$, $b\bar{b}$, $c\bar{c}$ and $q\bar{q}$ (2$^{\rm nd}$ column), as suggested by the results of Section \ref{sec:FermiBubbles}, and we indicate in parenthesis the $\chi^2_{\rm min}$ value obtained in the fit of the Fermi bubbles excess. In the last column we put the value of the ratio $s/\Lambda^2$; this value estimates the goodness of the effective field theory approach [see text, Eq.~(\ref{eq:estimator})]. {\underline{Color key}}. The color of each line is related to the strongest tension observed in the phenomenological analysis.  Operators marked in red can not fit the Fermi bubbles excess; operators in purple are ruled out by DD experiments; operators in yellow do not give a reliable effective field theory description; operators in grey do not reproduce the observed amount of relic abundance.}}
\label{tab:FermionDMproprieties}
\end{table}

Analyzing the operators one by one, we base our study on the following four criteria.
\begin{enumerate}

\item  We compute the gamma-ray flux originated from DM annihilation according to Eq.~(\ref{eq:Flux}). The analyzed operator must reproduce the Fermi bubbles excess,  as described in Section~\ref{sec:FermiBubbles}.

\item We compute the DM relic density solving numerically the Boltzmann equation, as summarized in Appendix~\ref{App:Boltzmann}, and we request each operator to reproduce the value observed by the Planck collaboration
\cite{Ade:2013zuv}.
 Note that if the value of the DM coupling corresponding to the correct relic density turns out to be larger than the one from the fit of the Fermi bubbles excess, then we will end up with too many photons from DM annihilation, thus contradicting the gamma ray data.
If, on the other hand, the coupling is smaller, then the explanation of the Fermi bubbles excess can not be realized.

\item For each operator, we impose the stringent bound on the DM-nucleon spin-independent (SI) elastic cross section from the null result of the XENON100 experiment  \cite{Aprile:2012nq}. Spin-dependent (SD) interactions are  constrained by PICASSO \cite{BarnabeHeider:2005pg}, SIMPLE \cite{Felizardo:2011uw},
ZEPLIN-II \cite{al.:2007xs}, and XENON100 \cite{Aprile:2013doa} data but the corresponding bounds are in general weaker w.r.t. the SI ones. Nevertheless, we will comment on the role of SD DM searches in the context of our analysis.

\item Finally, we investigate the validity of the effective field theory description.
We compare the best-fit value of the DM mass, obtained from the analysis of the Fermi bubbles excess,
 with a naive estimation of the cut-off scale, which is inversely related to the best-fit value of the
 coupling constant.

\end{enumerate}

 We summarize in Table~\ref{tab:FermionDMproprieties} the main properties of the effective operators in Eqs.~(\ref{eq:Scalar}-\ref{eq:Tensor}) associated with the DM annihilation cross section (relevant for the computation of the relic abundance and the fit of the Fermi bubbles excess), and the SI elastic cross section on nuclei relevant for DD. The explicit formulae are given in Appendix \ref{app:A}. Moreover, the color code (see the caption in Table~\ref{tab:FermionDMproprieties}) is used to indicate the main reason why the corresponding operator with a certain final state fails; for instance, the tensor operator with the $b\bar{b}$ final state can simultaneously explain the Fermi bubbles excess and reproduce the correct DM relic density as shown in the right panel of Fig.~\ref{fig:VectorTensorOperators}, but it needs a large coupling constant (small cut-off scale) due to the weak magnetic dipole moment interaction, hence voiding the validity of the effective field theory description. According to our color code, we mark this operator in yellow.

 Our main results are presented in the plane DM mass-DM coupling, $[M_{\rm DM}, G^f]$; in Fig.~\ref{fig:PseudoscalarPseudovectorOperators} and in Fig.~\ref{fig:VectorTensorOperators} we show the chi-square fit of the Fermi bubbles excess together with the contour reproducing the correct DM relic density, and in Fig.~\ref{fig:XENON100} we include the latest XENON100 constraints~\cite{Aprile:2012nq}.

 Before discussing the results, let us notice that
 the scalar, pseudoscalar and pseudovector operators in Eqs.~(\ref{eq:Scalar},\ref{eq:Pseudoscalar},\ref{eq:Pseudovector}) are valid considering both Dirac and Majorana DM, while the vector and tensor structure in Eqs.~(\ref{eq:Vector},\ref{eq:Tensor}) can be written only considering Dirac DM. From the phenomenological point of view adopted in our analysis, the difference between Majorana and Dirac DM can be described as follows.
 First, from Eq.~\ref{eq:Flux}, there exists a factor 2 of difference in the photon flux for a given value of $\langle\sigma v \rangle_f$. In addition, the cross section for Majorana DM is bigger w.r.t. the Dirac case given the same coupling constant.\footnote{Roughly speaking, the reason is that in the Majorana case the same diagram enters twice in the computation of the amplitude, because of the self-conjugate nature of the external legs \cite{Ciafaloni:2011sa}.}
 As a result, going from Dirac to Majorana DM, the confidence region obtained from the fit of
  the Fermi bubbles excess will shift downward in the plane
 $[M_{\rm DM}, G^f]$. The similar behavior occurs to the computation of the DM
 relic density (see Appendix~\ref{App:Boltzmann} for details) so that the relative position
  of the confidence region obtained from the fit w.r.t. the contour reproducing the DM relic density is the same for Majorana and Dirac DM.

 We summarize the results of our analysis as follows.

\begin{itemize}
\item
{\underline{Scalar operator $\mathcal{O}_{\rm S}^{f}$}}. The annihilation cross section is mass and velocity suppressed. As a consequence, due to the small DM velocity today ($v\sim 10^{-3}$), this type of operators is not able to produce the Fermi bubbles excess, and henceforth it will be neglected.

\item
{\underline{Pseudoscalar operator $\mathcal{O}_{\rm PS}^{f}$}}.
We start our discussion with Dirac DM.
The annihilation cross section is mass suppressed but not velocity suppressed, thus excluding immediately light quarks in the final state, for both the CP-preserving ($G_{\rm PSA}^{f}$) and CP-violating ($G_{\rm PS}^f$) operators [see Eq. (\ref{eq:xsecPS})]. In the following, we set for simplicity $G_{\rm PSA}^{f}=0$; we have checked that our conclusions remain intact in the opposite case, $G_{\rm PS}^{f}=0$.\\
Final states involving $\tau^+\tau^-$, $c\bar{c}$, $b\bar{b}$ provide a good fit to the Fermi bubbles excess, as shown in the left panel of Fig.~\ref{fig:PseudoscalarPseudovectorOperators}, where we plot the 68~\% and 95~\% confidence regions.  Only the confidence region obtained considering the $b\bar{b}$ final state leads to the correct DM relic density while $c\bar{c}$ and $\tau^+\tau^-$ channels do not have overlap.\\
In terms of DD searches, in the non-relativistic limit, the operator of $G_{\rm PS}^f$ ($G_{\rm PSA}^f$) has SI (SD) interactions~\cite{Fan:2010gt}; the leading order term of the DM-nucleus scattering is proportional to $\vec{S}_\chi \cdot  \vec{q}$ ($\vec{S}_\chi \cdot  \vec{q}~\vec{S}_{\rm N} \cdot  \vec{q}$), where $\vec{S}_\chi (\vec{S}_{\rm N})$ is the spin of the DM particle (nucleus) and $\vec{q}$ is the transferred momentum, typically of $\mathcal{O(\rm{MeV})}$ for DD experiments. Therefore, the pseudoscalar operator can not be constrained by direct search experiments because of the large momentum suppression~\cite{Cheung:2012gi}.\\
Finally, it is worthwhile to comment on the reliability of the effective field theory approach. Considering, for example, the $b\bar{b}$ final state, the best-fit value of the pseudoscalar coupling $G_{\rm PS}^{b}\sim 6\cdot 10^{-7}$ GeV$^{-3}$ leads to a naive estimation of the cut-off $\Lambda\sim (G_{\rm PS}^{b})^{-1/3} \sim 120$ GeV. On the other hand, with the corresponding best-fit value of the DM mass $M_{\chi}\sim 57$ GeV, we argue that the effective field theory approach based on the assumptions enumerated in Section \ref{sec:EFTapproach} might not be trustable in this case. The quantitative estimate of the validity of the effective approach can be characterized by the ratio $s/\Lambda^2$, resulting from the expansion,
\begin{equation}\label{eq:estimator}
\frac{1}{s-\Lambda^2}=-\frac{1}{\Lambda^2}-\frac{s}{\Lambda^4}+\mathcal{O}(s^2)~,
\end{equation}
where $\sqrt{s} \simeq 2M_{\chi}$.\footnote{Notice that in our naive estimation of the cut-off, i.e. $G^f_{\rm PS}=c^f_{\rm PS}/\Lambda^3$, we assume $c^f_{\rm PS}\sim \mathcal{O}(1)$. In the following, we will always adopt this assumption. However, it is straightforward to translate our results into general cases of $c^f_{\rm PS} \neq 1.$}  This parameter corresponds to the difference between the zeroth and first order approximation in the limit of $s \ll \Lambda$. We present these values in the last column of Table~\ref{tab:FermionDMproprieties}.\footnote{It is clear from Table~\ref{tab:FermionDMproprieties} that $\tau$ always has small $s/\Lambda^2$ in that $\tau$ produces the hardest FSR photon spectrum that makes smaller both best-fit $M_\chi$ and $\sigma v$. It in turn implies larger $\Lambda$ and smaller $s/\Lambda^2 (\sim 4M_\chi^2/\Lambda^2)$.} We consider the effective operator approach as a good approximation whenever $s/\Lambda^2< 10~\%$. This criterium disfavors the effective theory description of annihilations into $c\bar{c}$ ($43.7~\%$) and aforementioned $b\bar{b}$ ($78.5~\%$). As argued before, for Majorana DM both the confidence region obtained from the fit of the Fermi bubbles excess and the contour reproducing the correct DM relic density will shift downward with their relative position fixed, thus resulting in a smaller coupling constant and a bigger cutoff scale. Therefore, the reliability of the effective approach will improve for Majorana DM. For the $b\overline{b}$ final state, however, $s/\Lambda^2\sim 39.2~\%$, is still too large to validate the effective field theory description.\\
To sum, the effective field theory description of DM annihilation based on a pseudoscalar operator fails to explain the Fermi bubbles excess.

\item
{\underline{Vector operator $\mathcal{O}_{\rm V}^{f}$}}. It is well known that the annihilation cross section for the vector operator is characterized by an unsuppressed s-wave component [see Eq.~(\ref{eq:xsecV})], providing a natural realization of the WIMP miracle paradigm. We start with the case $G_{\rm VA}^{f}=0$ and present results in the right panel of Fig.~\ref{fig:PseudoscalarPseudovectorOperators}. Since the DM annihilation cross section does not depend on the final state quark mass in the limit  $M_\chi \gg m_f$, we have only one single orange line reproducing the correct DM relic density for the $b\bar{b}$, $c\bar{c}$ and $q\bar{q}$ channels; the green line refers to the $\tau^+\tau^-$ final state, which does not carry the ${\rm SU}(3)$ color charge. We find that only the $b\overline{b}$ final state can reproduce both the Fermi bubbles excess and the correct DM relic density. Additionally, for this channel, the best fit value of the coupling $G_{\rm V}^{b}\sim 2\cdot 10^{-6}$ GeV$^{-2}$ leads to $\Lambda\sim (G_{\rm V}^{b})^{-1/2}\sim 760$ GeV for the cutoff scale with $s/\Lambda^2\sim 1.9~\%$. This is also the case for $c\bar{c}$, $q\bar{q}$ and $\tau^+\tau^-$ channels in the wake of the unsuppressed s-wave contribution.

  We now switch to DM-nucleus scattering relevant for DD experiments. A vector operator has a non-zero SI elastic cross section on nuclei  \cite{Cheung:2012gi}; the null results of the XENON100 experiment \cite{Aprile:2012nq}, therefore,  can put strong bounds on the $[M_{\chi}, G_{\rm V}^{f}]$ parameter space. In particular, we find that the elastic cross section on nuclei mediated by the interactions of DM with the valence quarks (q=u,d) of the nucleon  is ruled out by several orders of magnitude considering the XENON100 bound. On the other hand, sea quarks do not contribute at the tree level to the nuclear vector current, due to the exact cancellation between particles and antiparticles. It could lead to a superficial conclusion that the $b\bar{b}$, $c\bar{c}$, and $\tau^+\tau^-$ final state are not constrained at all by the XENON100 experiment. These final states, nevertheless, can have a sizable $\sigma_{\rm SI}$ at one-loop through electromagnetic interactions: the photon emitted from virtual fermion loop couples to the nucleus of charge $Z$ (see Fig.~\ref{fig:LoopDiagram}).  The importance of these interactions has been already emphasized in Ref.~\cite{Fox:2011fx,Kopp:2009et} in the context of leptophillic DM (see also Ref.~\cite{Haisch:2013uaa}). In Appendix \ref{app:DDoneLoop}, we review the one-loop computation for channels of interest. The resulting DD bounds are presented in Fig.~\ref{fig:XENON100}. Strikingly, the XENON100 null result is able to exclude a relatively large region of the parameter space even in the case of the loop-induced processes. As a result, the $c\bar{c}$ interpretation of the Fermi bubbles excess is in strong tension with the DD experimental bound (Fig.~\ref{fig:XENON100}, central panel). The best fit region for the $b\bar{b}$ final state lies in the allowed region (Fig.~\ref{fig:XENON100}, left panel) but well within the reach of future sensitivity \cite{Aprile:2012zx}. Annihilation into $\tau^+\tau^-$ final state, on the contrary, can fit the Fermi bubbles signal with light DM masses, a region where the XENON100 bound becomes weaker.

For the opposite situation, namely $G_{\rm V}^{f}=0$, $G_{\rm VA}^{f}\neq 0$, the annihilation cross section has unsuppressed s-wave contribution, and consequently we obtain quite similar results. In particular, annihilation into $b\overline{b}$ exhibits an overlap between the Fermi bubbles excess's confidence region and the contour reproducing the correct DM  relic density. The most important difference comes from the DD constraint. At the tree-level, the operator $(\bar{\chi}\gamma^{\mu}\chi)(\bar{f}\gamma_{\mu}\gamma^{5}f)$ is velocity suppressed~\cite{Fan:2010gt, Cheung:2012gi} and has only spin dependent interactions with nuclei. Hence it can avoid the stringent XENON100 SI bound.\footnote{Considering
  the operator $(\bar{\chi}\gamma^{\mu}\chi)(\bar{q}\gamma_{\mu}\gamma^{5}q)$ with quarks in the final state, one may be worried about the contribution to DD searches from typical anomaly triangle loop diagrams associated with two gluons. Integrating out the loop in the heavy quark limit, we find that this process is equivalent to the effective interaction
    \begin{equation}\label{eq:effectivegluons}
\mathcal{L}_{eff}=-\frac{iG_{\rm A}^{q}g_s^2}{48\sqrt{2}\pi^2m_q^2}(\bar{\chi}\gamma^{\mu}\chi)
[G^{\rho\sigma}(\partial_{\sigma}\tilde{G}_{\rho\mu})]~,
  \end{equation}
  where $g_s$ is the strong coupling constant and $\tilde{G}_{\rho\mu}$ is the dual field strength $\tilde{G}_{ab}=\epsilon_{abcd}G^{cd}/2$. Because of the C- and P-preserving properties of QCD, the effective operator in Eq. (\ref{eq:effectivegluons}) inherits the DD properties from $(\bar{\chi}\gamma^{\mu}\chi)(\bar{f}\gamma_{\mu}\gamma^{5}f)$, resulting in a velocity suppressed SD cross section.
  }

  In summary, we argue that DM annihilation into $b\overline{b}$ through the vector operator in Eq.~(\ref{eq:Vector}) is a forefront candidate to explain the Fermi bubbles excess in the context of the effective approach.

\item
{\underline{Pseudovector operator $\mathcal{O}_{\rm PV}^{f}$}}.

The annihilation cross section for the pseudovector operator has a s-wave contribution only in the presence of
a non-zero axial-vector coupling $G_{\rm PVA}^{f}$ [see Eqs.~(\ref{eq:Pseudovector}, \ref{eq:xsecPV})].
As a result, we simply set in our analysis $G_{\rm PV}^{f}=0$. The s-wave is mass-suppressed due to the chirality flip, while
the p-wave does not have mass-suppression.\footnote{
Let us quickly recap the origin of the mass-suppressed
s-wave for the pseudovector operator in Eq.~(\ref{eq:Pseudovector}).
Considering the DM side, $\bar{\chi}\gamma^{0}\gamma^{5}\chi$ is the
only component contributing to the s-wave annihilation, as can be inferred taking the non-relativistic limit \cite{Cheung:2012gi}. In addition, considering the SM side, the combination $\bar{f}\gamma_{0}f$ is zero for a fermion-antifermion system;
therefore the only possible non-zero contraction of $\bar{\chi}\gamma^{0}\gamma^{5}\chi$ involves $\bar{f}\gamma_{0}\gamma^5 f$.
Moreover, from the CP transformation property
$\bar{\psi}\gamma^{\mu}\gamma^5\psi \overset{\rm CP}{\Longrightarrow} (-1)^{\mu}\bar{\psi}\gamma^{\mu}\gamma^5\psi$, it follows that both
$\bar{\chi}\gamma^{0}\gamma^{5}\chi$ and $\bar{f}\gamma_{0}\gamma^5 f$ have $\mathrm{CP} =-1$;
this in turn implies that the s-wave annihilation must have total spin $S=0$ \cite{Kumar:2013iva} because of ${\rm CP}=(-1)^{{\rm S}+1}$. As a consequence, the condition $S=0$ on $\bar{f}\gamma_{0}\gamma^5 f$  forces the SM
fermion-antifermion pair to carry the same helicity, implying the presence of a chirality flip (mass insertion).
 Following the same logic, the chirality flip is not required for the p-wave annihilation.}
This fact rules out immediately the light quark final states as explanation of the Fermi bubbles excess.
Annihilations into $\tau^+\tau^-$, $b\bar{b}$ and $c\bar{c}$, on the contrary, can fit the data as shown in the left
panel of Fig.~\ref{fig:VectorTensorOperators}.
As far as the computation of the relic density is concerned, only DM annihilation
into $\tau^+\tau^-$ can reproduce the observed value. Notice that this result completely overturns the scenario
outlined in Fig.~\ref{fig:BubbleFit}. The reason is that for the pseudovector operator the p-wave becomes
dominant at freeze-out epoch; a naive estimation based on Eq.~(\ref{eq:xsecPV}), in fact, shows that
$\frac{{\rm s-wave}}{{\rm p-wave}}\sim m_f^2/v^2M_{\chi}^2$. During the freeze-out, i.e. $v^2\sim 1/4$, the p-wave
dominates over the s-wave for all the final state fermions considered in our analysis.
As mentioned before, this fact  singles out the $\tau^+ \tau^-$ final state as the only one able to
reproduce both the Fermi bubbles excess and the correct value of relic density.
\\
%
Moreover -- analyzing the reliability of the effective field theory approach -- we find that the $b\bar{b}$ and $c\bar{c}$ final states have a relatively large $s/\Lambda^2$, while, for $\tau^+\tau^-$, the effective theory provides a good description.\\
\\
Unlike the vector operator, the pseudovector one is non-zero for Majorana DM.
As argued before, for Majorana DM both the confidence region obtained from the fit of the Fermi bubbles excess and the contour reproducing the correct DM relic density will shift downward, keeping their relative position fixed.
 As a result, the $\tau^+\tau^-$ channel can
realize the Fermi bubbles excess while $b\bar{b}$ and $c\bar{c}$ can not reproduce the correct value of relic density; the only difference is that the $b\bar{b}$ and $c\bar{c}$ channels have a smaller $s/\Lambda^2$ compared to the Dirac case, $5~\%$ for $c\bar{c}$ and $12~\%$ for $b\bar{b}$.

   To summarize, DM (either Majorana or Dirac) annihilation into $\tau^+\tau^-$ through the pseudovector operator in Eq.~(\ref{eq:Pseudovector}) is a forefront candidate to explain the Fermi bubbles excess.

\item
{\underline{Tensor operator $\mathcal{O}_{\rm T}^{f}$}}. The tensor structure in Eq.~(\ref{eq:Tensor}) involves the chirality flip, and therefore leads to a mass-suppressed s-wave contribution in the annihilation cross section, as shown in Eq.~(\ref{eq:xsecT}), excluding the light quark channels. The same behavior is shared both by the CP-preserving ($G_{\rm T}^{f}$) and CP-violating
 ($G_{\rm TA}^f$) contribution [see Eq. (\ref{eq:xsecT})]. In the following, we simply set $G_{\rm TA}^{f}=0$; we have checked that our conclusions keep unchanged in the opposite case. The results are very similar to those of the pseudoscalar case. We plot our results in the right panel of Fig.~\ref{fig:VectorTensorOperators}.\\
 Final states involving $\tau^+\tau^-$, $c\bar{c}$, $b\bar{b}$ provide a good fit to the Fermi bubbles excess;
among them, annihilation into $b\bar{b}$ is the only channel leading to the correct amount of relic density but it suffers from the poor reliability of the effective description as a result of the weak dipole moment interactions.\\
To sum up, the tensor operator does not provide the explanation to the Fermi bubbles excess in the context of the effective approach.
\end{itemize}
\begin{figure}[!htb!]
   \begin{minipage}{0.4\textwidth}
   \centering
   \includegraphics[scale=0.65]{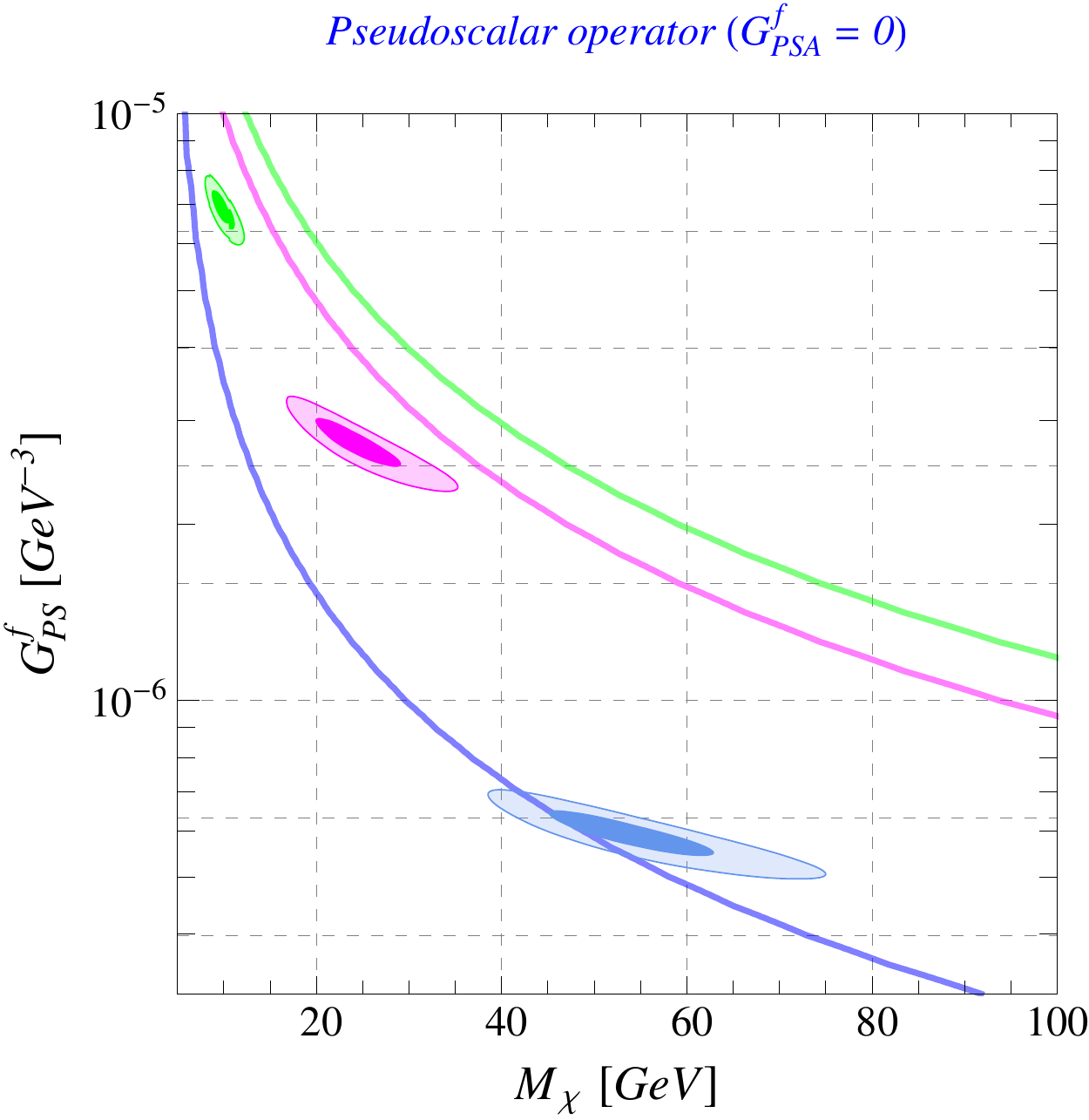}
    \end{minipage}\hspace{2 cm}
   \begin{minipage}{0.4\textwidth}
    \centering
    \includegraphics[scale=0.65]{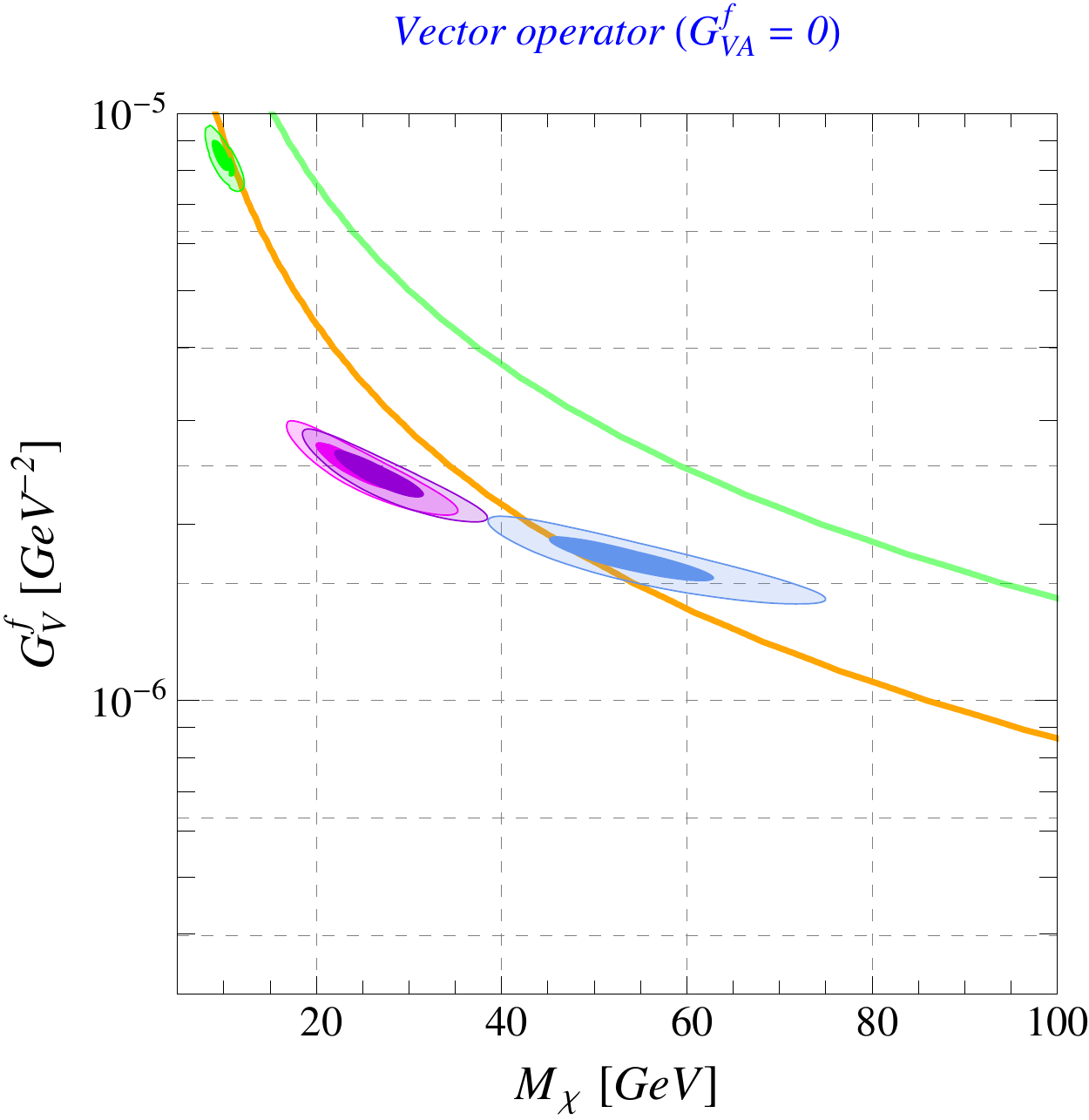}
    \end{minipage}
 \caption{\emph{Chi-square results (68~\% C.L., darker region; 99~\% C.L., lighter region)  for the fit of the Fermi bubbles excess
  in the plane $[M_{\chi}, G_{i}^{f}]$. The color code follows Fig.~\ref{fig:BubbleFit}.
  We superimpose the contours reproducing the correct DM relic density. For the vector operator, the DM relic density calculation does not depend on the mass of the final state fermions, and thus we plot one single orange line for the $b\bar{b}$, $c\bar{c}$ and $q\bar{q}$ channels. The $\tau^+\tau^-$ channel differs because of the absence of the color factor.
  {\underline{Left panel}}: Pseudoscalar operator.  {\underline{Right panel}}: Vector operator.}}
 \label{fig:PseudoscalarPseudovectorOperators}
 \end{figure}
 \begin{figure}[!htb!]
  \begin{minipage}{0.4\textwidth}
   \centering
   \includegraphics[scale=0.65]{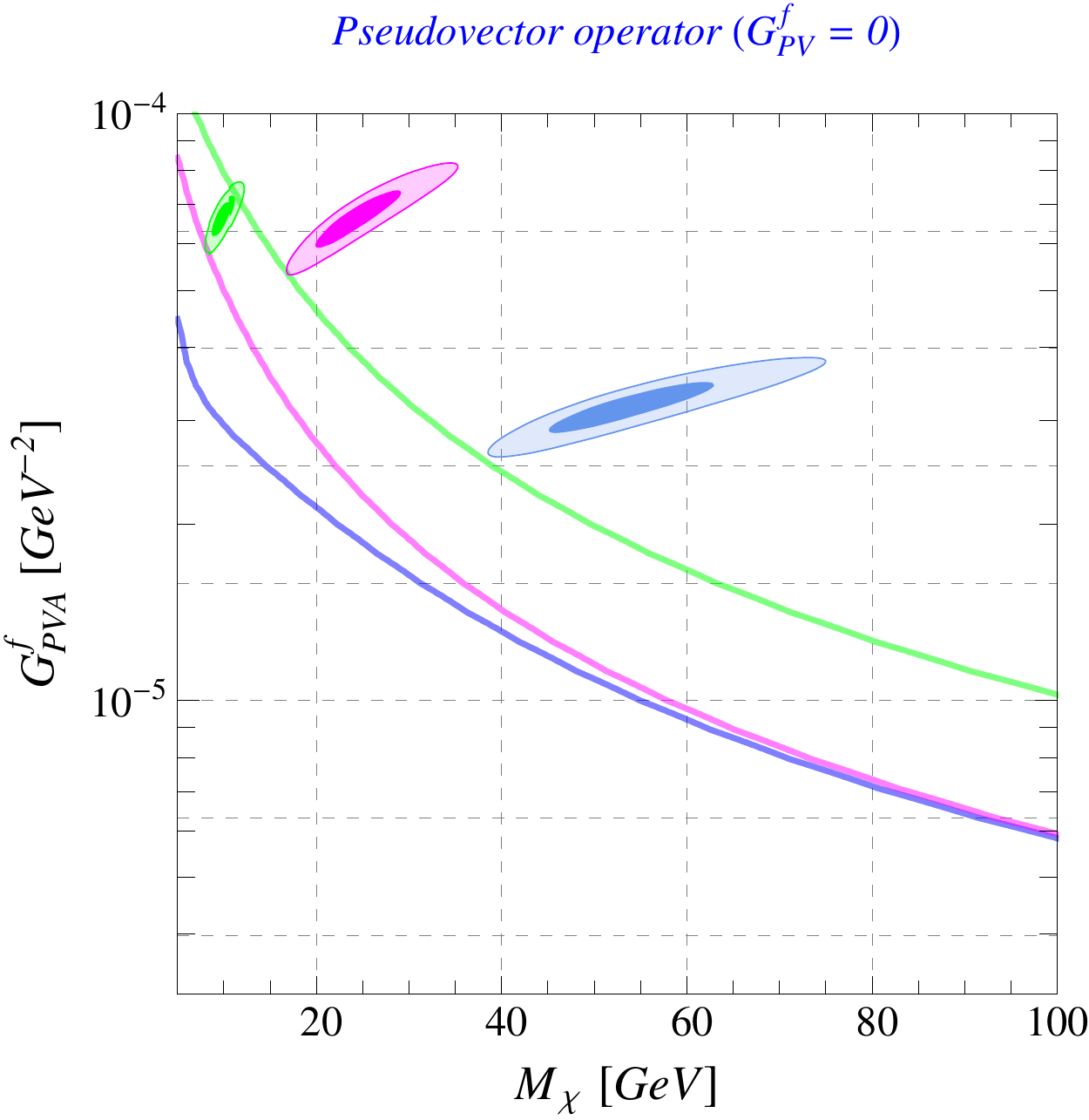}
    \end{minipage}\hspace{2 cm}
   \begin{minipage}{0.4\textwidth}
    \centering
    \includegraphics[scale=0.65]{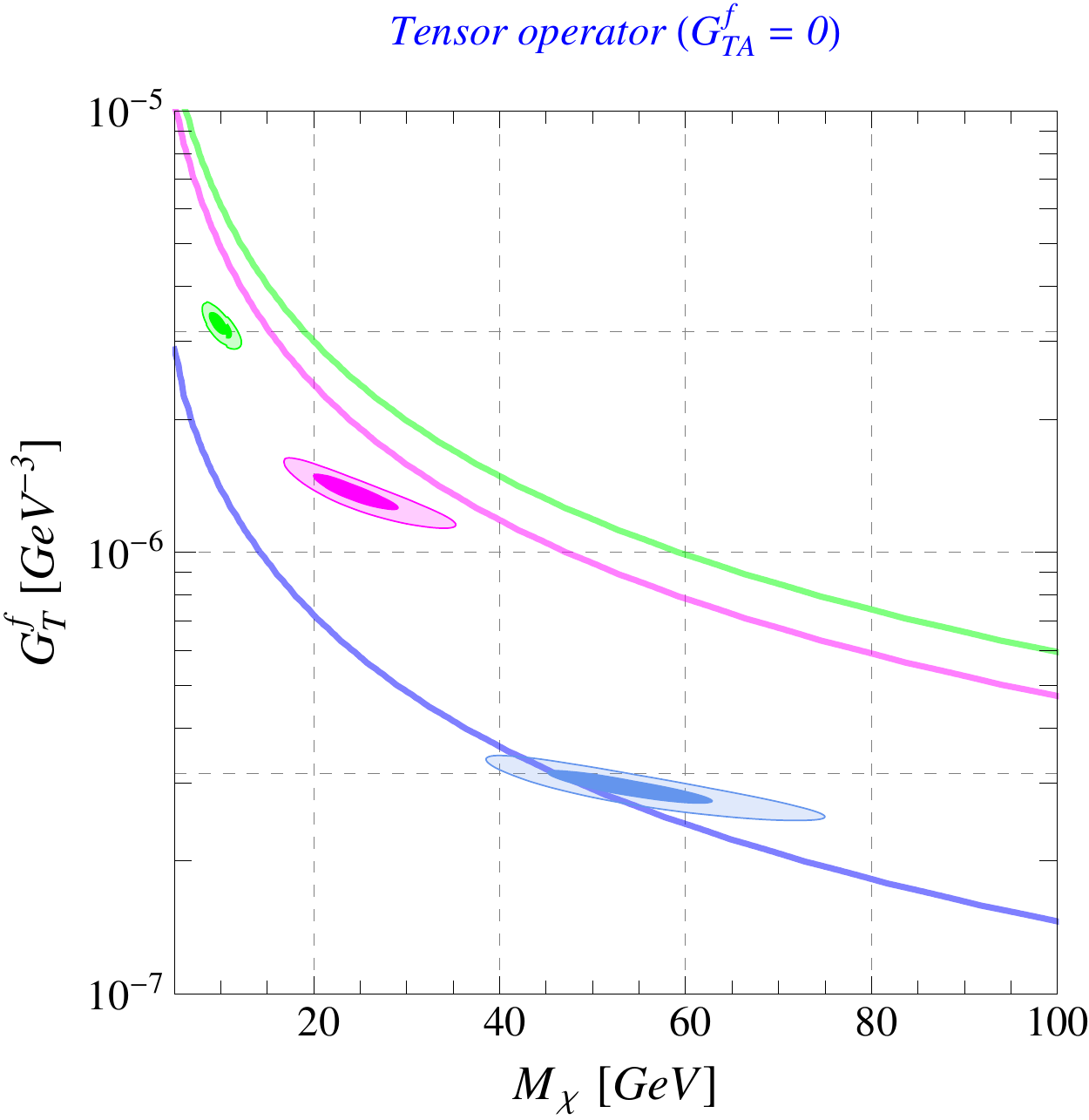}
    \end{minipage}
 \caption{\emph{The same as in Fig.~\ref{fig:PseudoscalarPseudovectorOperators}. {\underline{Left panel}}: Pseudovector operator.  {\underline{Right panel}}: Tensor operator.}}
 \label{fig:VectorTensorOperators}
\end{figure}
All in all, the analysis of the Fermi bubbles excess through the effective operators listed in Eqs.~(\ref{eq:Scalar}-\ref{eq:Tensor}) indicates some interesting scenarios.
If the DM interactions with SM fermions can be described by one single effective operator, the request to reproduce the correct relic density and avoid the bound from the XENON100 experiment allows only two cases: \textit{i}) DM annihilation into $b\overline{b}$ through the vector operator in Eq.~(\ref{eq:Vector}), with mass $M_{\chi} \simeq 52$ GeV or \textit{ii}) DM annihilation (either Majorana or Dirac) into $\tau^+\tau^-$ through the pseudovector operator in Eq.~(\ref{eq:Pseudovector}), with mass $M_{\chi} \simeq 10$ GeV. The latter scenario relies on the interplay between the s-wave (dominant for the annihilation of DM today, i.e. for the computation of the gamma-ray flux) and the p-wave (compatible in magnitude with the s-wave during the freeze-out epoch).
\begin{figure}[!htb]
\minipage{0.32\textwidth}
  \includegraphics[width=\linewidth]{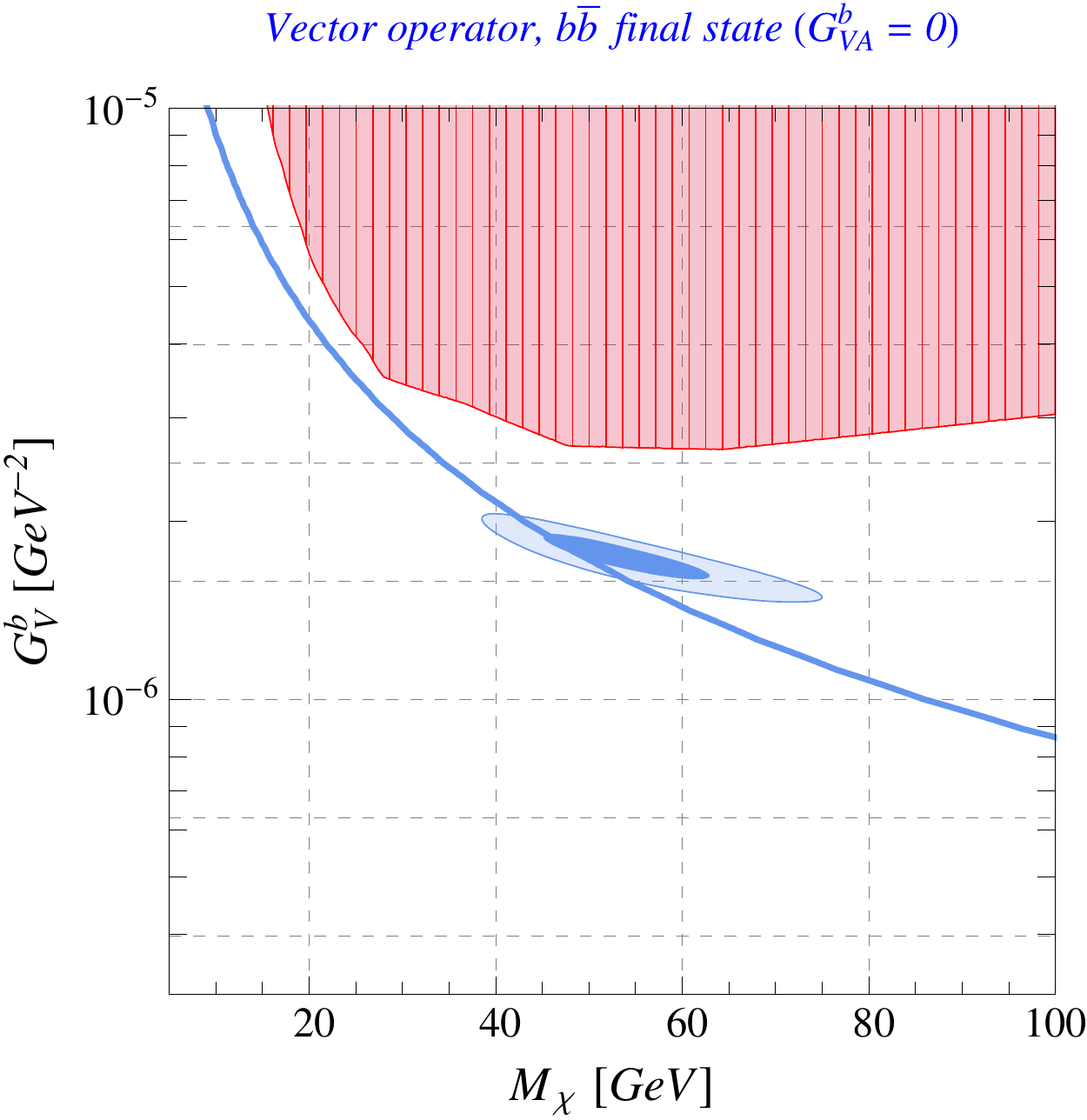}
\endminipage\hfill
\minipage{0.32\textwidth}
  \includegraphics[width=\linewidth]{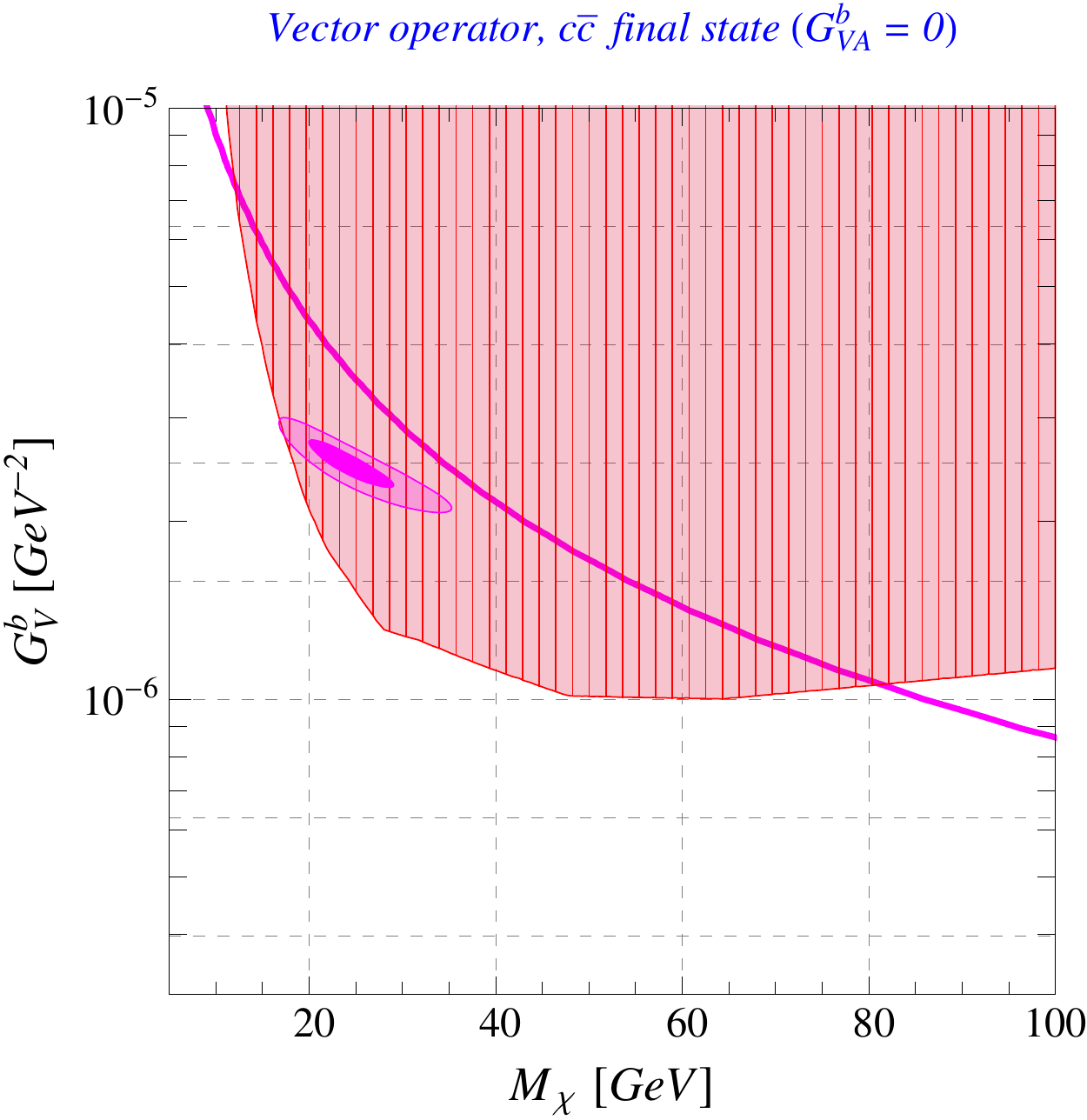}
\endminipage\hfill
\minipage{0.32\textwidth}
  \includegraphics[width=\linewidth]{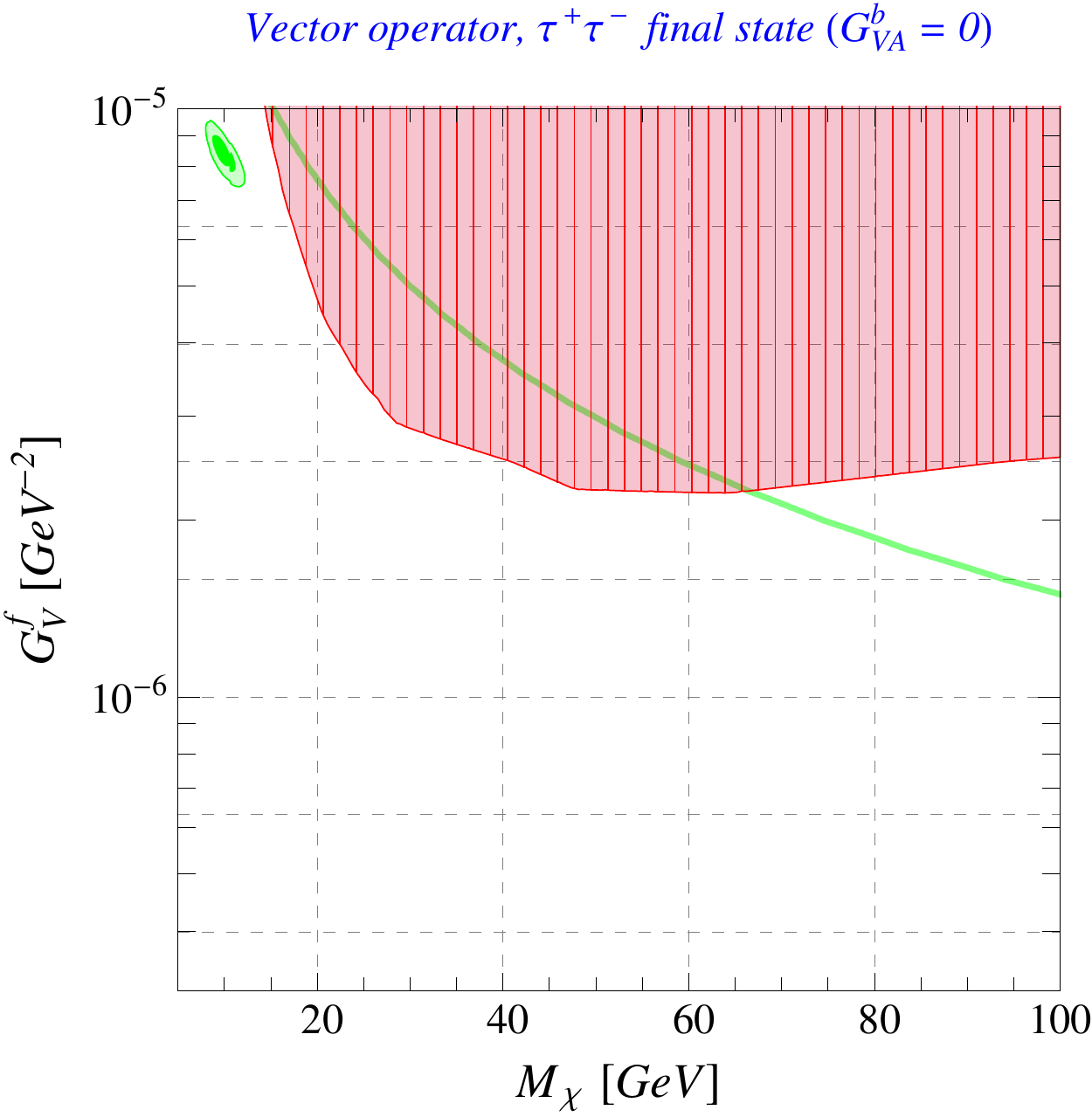}
\endminipage
 \caption{\textit{Region of the parameter space $[M_{\chi}, G_{\rm V}^f]$ excluded at 90~\% C.L. by the XENON100
 experiment \cite{Aprile:2012nq} (red region). We show $b\overline{b}$ final state (left panel), $c\overline{c}$ final state (central panel) and $\tau^+\tau^-$ final state (right panel) for the vector operator in Eq.~(\ref{eq:Vector}). Elastic cross sections are evaluated at one-loop level (see text for details, and Appendix~\ref{app:DDoneLoop} for the analytical expressions). For each case, we also show the confidence region obtained from the fit of the Fermi bubbles excess, and the contour reproducing the correct value of relic density (see the right panel of Fig.~\ref{fig:PseudoscalarPseudovectorOperators} and the corresponding caption).}}
 \label{fig:XENON100}
\end{figure}

\section{Towards Concrete Models}
\label{sec:MSSM}
The effective field theory description, discussed in Section~\ref{sec:EFTapproach}, is based on a set of assumptions that limit the variety of physical phenomena that can be captured. The purpose of this Section is to bridge part of this gap studying two concrete models that do not fall into the territory of the effective field theory approach. In Section~\ref{sec:MSSM_Higgs}, we study the scalar Higgs portal,\footnote{
{\bf Note added:} While this work was in its final stages, Ref.~\cite{Okada:2013bna} appeared. In this paper, the Higgs portal is proposed as a possible explanation of the Fermi bubbles excess.} while in Section~\ref{sec:MSSM_Zprime} we propose a toy-model with fermionic DM where the mediator between the dark and the visible sector is a light leptophillic $Z^{\prime}$ gauge boson.

Before proceeding, let us briefly consider some other possibilities. First,
notice that a model based on the $Z$ exchange either is excluded by the
XENON100 constraint (assuming a vector-like $Z$-DM coupling)
or can not yield the correct DM relic density (assuming a pseudovector-like $Z$-DM coupling). Another intriguing possibility is to realize the Fermi bubbles excess in the context of a supersymmetric model.
A light Bino-like neutralino~\cite{Boehm:2013qva,Hagiwara:2013qya,Pierce:2013rda},
annihilating into $\tau^+\tau^-$ via the t-channel exchange of a light stau, is a
potential candidate. At zero velocity,
we have $\sigma v \simeq (g_L g_R)^2 m_\chi^2 / 8 \pi m_{\tilde\tau}^4 \sim 5.6 \times
10^{-27} {\rm cm}^3{\rm s}^{-1}  \times ( 76~{\rm GeV}/ m_{\tilde \tau})^4$. As noticed in Ref.~\cite{Hagiwara:2013qya}, however, the light Bino-like neutralino
can not have the right relic density without an additional entropy injection. Despite this daunting conclusion, it is still an
interesting open question if supersymmetric models can
accommodate the
Fermi bubbles excess, and we will leave it to future
study. Finally, in Ref.~\cite{Anchordoqui:2013pta} the Fermi bubbles excess has been studied
in the context of the minimal hidden sector model  recently proposed in Ref.~\cite{Weinberg:2013kea}.

\subsection{Higgs exchange}
\label{sec:MSSM_Higgs}

In this Section, we add to the SM Lagrangian the following terms \cite{Silveira:1985rk,McDonald:1993ex,Burgess:2000yq}
\begin{equation}\label{eq:SSLagrangian}
\mathcal{L}_{S} = \frac{1}{2}(\partial_{\mu}S)(\partial^{\mu}S)
- \frac{1}{2}\mu_{S}^2 S^2 - \frac{1}{2}\lambda S^2|H|^2~,
\end{equation}
where $S$ is a gauge-singlet real scalar with mass, after electroweak symmetry breaking, given by $m_{S}=(\mu_{S}^2+\lambda v^2/2)^{1/2}$.
$H$ is the SM Higgs doublet with the vacuum expectation value (vev) $\langle H\rangle = v/\sqrt{2}=174$ GeV.
The interaction term $\lambda S^2|H|^2$ is known in the literature as the \textit{Higgs portal} \cite{Patt:2006fw}.\footnote{
It is also possible to construct a similar Lagrangian involving a gauge singlet fermionic field $\chi$.
In this case the interaction term $\lambda S^2|H|^2$ in Eq.~(\ref{eq:SSLagrangian})
is replaced by the following two terms \cite{Kim:2008pp,Baek:2011aa,LopezHonorez:2012kv}
\begin{equation}\label{eq:FermionicPortal}
\frac{1}{\Lambda_1} \overline{\chi}\chi|H|^2 + \frac{i}{\Lambda_2} \overline{\chi}\gamma^5\chi|H|^2~.
\end{equation}
The first operator in Eq.~(\ref{eq:FermionicPortal}) leads to a velocity-suppressed annihilation cross section, and therefore it can not reproduce the Fermi bubbles excess. The CP-violating operator $\overline{\chi}\gamma^5\chi|H|^2$ in Eq.~(\ref{eq:FermionicPortal}), on the contrary, is characterized by an unsuppressed s-wave.
Given that in this Section we are interested in minimal renormalizable realizations of the interplay
between SM and the DM sector, we focus our attention on the Lagrangian in Eq.~(\ref{eq:SSLagrangian}). Another interesting possibility is to construct a vector Higgs portal through the interaction
\begin{equation}
\frac{\lambda_{\rm V}}{2}|H|^2V_{\mu}V^{\mu}~,
\end{equation}
where $V_{\mu}$ is a massive gauge boson, singlet under the SM gauge group. The resulting DM phenomenology turns out to be very similar to the singlet scalar case, as discussed in Refs.~ \cite{Lebedev:2011iq,Kanemura:2010sh,Baek:2012se}. As a consequence, we do not dedicate a separate discussion to this realization. }

Despite its simplicity, the phenomenological implications of the renormalizable interaction in Eq.~(\ref{eq:SSLagrangian}) are
 multiple and various \cite{Davoudiasl:2004be,Ham:2004cf, Patt:2006fw,O'Connell:2006wi,He:2007tt,Profumo:2007wc,Barger:2007im,He:2008qm,Ponton:2008zv,Lerner:2009xg,Farina:2009ez,Bandyopadhyay:2010cc,Barger:2010mc,Guo:2010hq,Espinosa:2011ax,Profumo:2010kp,Djouadi:2012zc,Mambrini:2012ue,Mambrini:2011ik,Goudelis:2009zz}, focusing in particular on the possibility that $S$ can play the role of cold DM in the Universe. Imposing an extra $Z_2$ symmetry to ensure the stability of $S$, the scalar singlet can annihilate into all the SM final states due to the s-channel exchange of the Higgs boson, which could reproduce the right DM abundance (see Refs.~\cite{Cheung:2012xb,Cline:2013gha} for updated analyses). To be more quantitative, the annihilation cross section times relative velocity of the scalar singlet takes the form
 \begin{equation}\label{eq:ScalarSingletXSection}
 \sigma_{\rm S}v = \frac{2}{\sqrt{s}}\left[
 \frac{\lambda^2v^2}{(s-m_h^2)^2 + \Gamma_{h,\rm S}^2 m_h^2}
 \right]\Gamma_{h}(\sqrt{s})~,
 \end{equation}
 where $\Gamma_{h}(\sqrt{s})$ is the off-shell decay width of the Higgs boson, including all the SM final states. We evaluate this function using the public code HDECAY \cite{Djouadi:1997yw}.\footnote{In this way, we include both the large QCD corrections on the $q\bar{q}$ final state and the three- and four-body decays of the Higgs into $W^{(*)}W^{*}$, $Z^{(*)}Z^{*}$.}
 In the denominator of the propagator, $\Gamma_{h,\rm S}$ is the value
 of the decay width of the Higgs boson at $m_h=125$ GeV, and it consists of two pieces: in addition to the SM contribution, $\Gamma_{\rm vis}=4.07$ MeV, we have to include the invisible decay of the Higgs into two DM particles, $ \Gamma_{\rm inv}(h\to SS)$,  that is kinematically allowed if $m_{S} < m_h/2$. Therefore, we have
 \begin{equation}\label{eq:InvisibleWidth}
 \Gamma_{h,\rm S} \equiv \Gamma_{\rm vis} +  \Gamma_{\rm inv}(h\to SS)~,~~~ \Gamma_{\rm inv}(h\to SS) = \frac{\lambda^2 v^2}{32\pi m_h}\sqrt{1-\frac{4m_{S}^2}{m_h^2}}~\vartheta(m_h - 2m_S)~.
 \end{equation}
The search for a non-zero invisible Higgs decay is currently under investigation at the LHC \cite{ATLAS_inv,CMS_inv}. Stringent upper bounds have been placed, thus allowing to constrain, through Eq.~(\ref{eq:InvisibleWidth}), the parameters $m_S$ and $\lambda$.

Equipped with Eq.~(\ref{eq:ScalarSingletXSection}), we can derive the thermally averaged cross section according to the following integral
\begin{equation}
\langle \sigma_{\rm S}v\rangle = \int_{4m_{\rm S}^2}^{\infty}
ds \frac{s\sqrt{s-4m_{\rm S}^2}K_{1}(\sqrt{s}/T)}{16 T m_{\rm S}^4 K_{2}^2(m_{\rm S}/T)}
\sigma_{\rm S}v ~,
 \end{equation}
where $T$ is the Universe temperature in question and $K_{\alpha}$ are the modified Bessel functions of the second kind. For
the computation of the relic density, we use this definition without any approximation, solving numerically the Boltzmann equation.
We briefly review the main points of this analysis in Appendix~\ref{App:Boltzmann}. For the computation of the photon flux, however, we use in Eq.~(\ref{eq:Flux}) the zero-velocity approximation $\langle \sigma_{\rm S}v\rangle =
 \left.\sigma_{\rm S}v\right|_{s=4m_{\rm S}^2}$.\footnote{In the computation of the photon flux, we use the energy spectra provided by Ref.~\cite{Cirelli:2010xx}, thus including also the effects of the three- and four-body decays $h\to V^{(*)}V^{*}$, $V=W,Z$ which are particularly relevant close to the Higgs resonance.}

 Finally, through the Higgs couplings with quarks, the scalar singlet may also have a sizable SI elastic cross section on nuclei, $\sigma_{\rm SI}$, severely constrained by the XENON100 experiment. In the limit of $q^2 \ll m_s^2$, where $q$ is the momentum transferred, it has a remarkably simple form \cite{Cline:2013gha}
 \begin{equation}
 \sigma_{\rm SI} = \frac{\lambda^2 f_{\rm N}^2}{4\pi}\frac{\mu^2 m_{\rm N}^2}{m_h^4 m_S^2}~,
 \end{equation}
where $\mu\equiv m_{\rm N}m_S/(m_{\rm N}+m_S)$ is the DM-nucleon reduced mass,
$m_{\rm N}=0.946$ GeV is the nucleon mass, and $f_{\rm N}=0.303$ is a numerical coefficient of the Higgs-nucleon coupling, which
is described in the Appendix~\ref{app:A}.

Before checking if $S$ as DM can explain the bubbles and avoid the experimental constraints mentioned above, we would like to point out that this simple theoretical setup should be thought as a minimal parametrization that can be used as a reference for more complicated models. For example, in the case of a complex singlet where both degrees of freedom are kinematically accessible,
 the relic density doubles. Even non-renormalizable derivative couplings between the singlet and the Higgs, e.g. $(\partial_{\mu}|H|^2S\partial_{\mu}S)/\Lambda^2$, can be constrained in a similar way. These couplings arise, for instance, in composite Higgs models where both the Higgs and the singlet are pseudo Nambu-Goldstone bosons of a broken global symmetry \cite{Gripaios:2009pe}, whose breaking is characterized by the scale $\Lambda$. In this case, for an on-shell Higgs, we can identify $\lambda \to m_h^2/2\Lambda^2$ \cite{Frigerio:2012uc}.

Introducing the Higgs invisible branching ratio
\begin{equation}\label{eq:InvBR}
{\rm BR}_{\rm inv}(m_S,\lambda) \equiv \frac{\Gamma_{\rm inv}(h\to SS)}{\Gamma_{\rm inv}(h\to SS) + \Gamma_{\rm vis}}~,
\end{equation}
we fit the Fermi bubbles excess together with the results of all the Higgs searches under investigation at the LHC \cite{Falkowski:2013dza}.\footnote{Performing the fit of the LHC data, we assume that the Higgs couplings with the electroweak gauge bosons and fermions are equal to their SM values. As a consequence, we have as a free parameter only, the invisible branching ratio in Eq.~(\ref{eq:InvBR}). In this situation, Ref.~\cite{Falkowski:2013dza} obtains that ${\rm BR}_{\rm inv}> 22~\%$ is excluded at 95~\% C.L.. It is important to keep in mind, however, that this bound can be relaxed allowing for deviations
of the one-loop Higgs couplings with photons and gluons. If so, in fact, one finds that ${\rm BR}_{\rm inv}> 50~\%$ is excluded at 95~\% C.L..} We show our results in Fig.~\ref{fig:scalarhiggsexchange}. The red region is excluded by XENON100, while the green line is the relic density contour; the
99~\% confidence region obtained from the combined fit of the Fermi bubbles excess and Higgs data is
displayed in blue.
\begin{figure}
   \centering
   \includegraphics[scale=0.8]{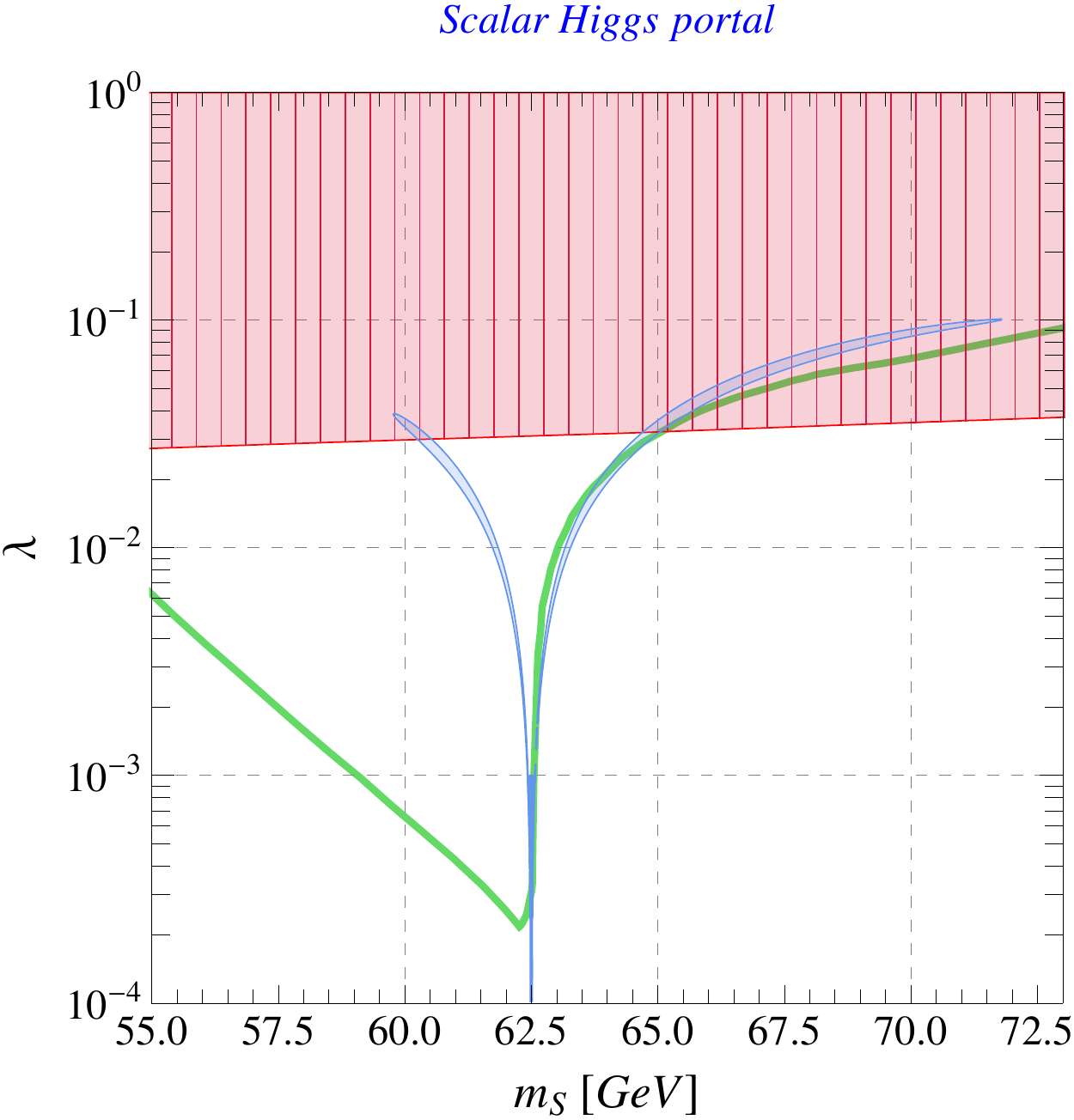}
 \caption{\emph{Chi-square result (99~\% C.L. contour, blue region) for the fit of the Fermi bubbles excess in the parameter space $[m_{\rm S}, \lambda]$ of the scalar Higgs portal model. We include in the fit also the LHC data constraining the invisible decay width of the Higgs, as discussed in Ref.~\cite{Falkowski:2013dza}. The red region is excluded at 90~\% C.L. by the XENON100 experiment, while the green line is the relic density contour.}}
 \label{fig:scalarhiggsexchange}
\end{figure}
This favored region retraces the Higgs resonance peaked at $m_{\rm S}=62.5$ GeV. The left-hand side,  $m_{\rm S}<62.5$, is cut around $m_{\rm S}\approx 60$ GeV; for lighter DM masses, in fact, the invisible decay of the Higgs
starts to be in conflict with the experimental data collected at the LHC, exceeding the allowed values. On the other hand, the region of $m_{\rm S}>62.5$ GeV does not suffer from the LHC bound. In the region of $62.5 \lesssim m_{\rm S}\lesssim 65$ GeV, the confidence region overlaps with the relic density contour while the region of $ m_{\rm S}\gtrsim 65$ GeV is excluded by the DD bounds.

The scalar Higgs portal, in conclusion, provides a simple, concrete and realistic DM model to explain the Fermi bubbles excess without in conflict with the DD SI bounds and the LHC Higgs data. In this context, the predicted value of the DM mass lies in the small window $62.5 \lesssim m_{\rm S}\lesssim 65$ GeV, a region that will be significantly reduced by the next generation of experiments (see Ref.~\cite{Cline:2013gha} for a detailed discussion).

\subsection{$Z^\prime$-exchange}
\label{sec:MSSM_Zprime}

In this Section, we study a toy model, where the $Z$ boson
mixes with an addition $U(1)_X$ gauge boson $Z^\prime$ and DM is charged under $U(1)_X$\footnote{The mixing could be,
for example, simply the kinetic mixing or a Stueckelberg extension of the SM~\cite{Kors:2004dx}.}. We employ the following assumptions.

\begin{enumerate}

\item Fermionic DM. From the effective field approach discussed above, the scalar DM annihilation
via $Z/Z^\prime$ is velocity suppressed, which cannot explain the Fermi bubbles excess.

\item Light $Z^\prime$. If $Z^\prime$ is heavy, our effective field theory approach applies. In addition,
the current LHC constraints~\cite{ATLAS:2013jma,CMS_zprime} imply $m_{Z^\prime}>$ few TeV or $m_{Z^\prime} \lesssim 100$ GeV.

\item Pseudovector DM-$Z^\prime$ interactions and leptophilic final states in order to evade the DD bound.

\item For simplicity, the couplings of $Z^\prime$ to SM fermions are proportional to those of $Z$ to SM fermions.

\end{enumerate}

As a result, the relevant Lagrangian for $Z^\prime$ can be written as
\be
\mathcal{L} \supset  g_{\chi \chi Z^\prime} \bar{\chi} \gamma^{\mu}\gamma^5 \chi Z^\prime_{\mu} +
g_{\ell \ell Z^\prime}  \bar{\ell} \gamma^{\mu} (g_{\ell_R} P_R +g_{\ell_L} P_L) \ell Z^\prime_{\mu},
\label{eq:DM_z'}
\ee
where $\ell$ refers to both charged leptons and neutrinos, $P_{L,R}$ are the projection operators, and $g_{\ell_{R,L}}$ are the SM lepton couplings to $Z$.
To reduce the number of free parameters, we set $g_{\chi \chi Z^\prime}=1$ and $m_Z^\prime=17.5$ GeV.\footnote{The lesson from the Higgs-exchange case implies the need of the resonance. Moreover, the leptonic final states, which will be dominated by $\tau^+\tau^-$, feature low-mass DM ($\sim 10$ GeV) as shown in Fig.~\ref{fig:BubbleFit}. Therefore, the mass of $Z^\prime$ is chosen around 20 GeV.}.
\begin{figure}[!htb!]
   \centering
   \includegraphics[scale=1]{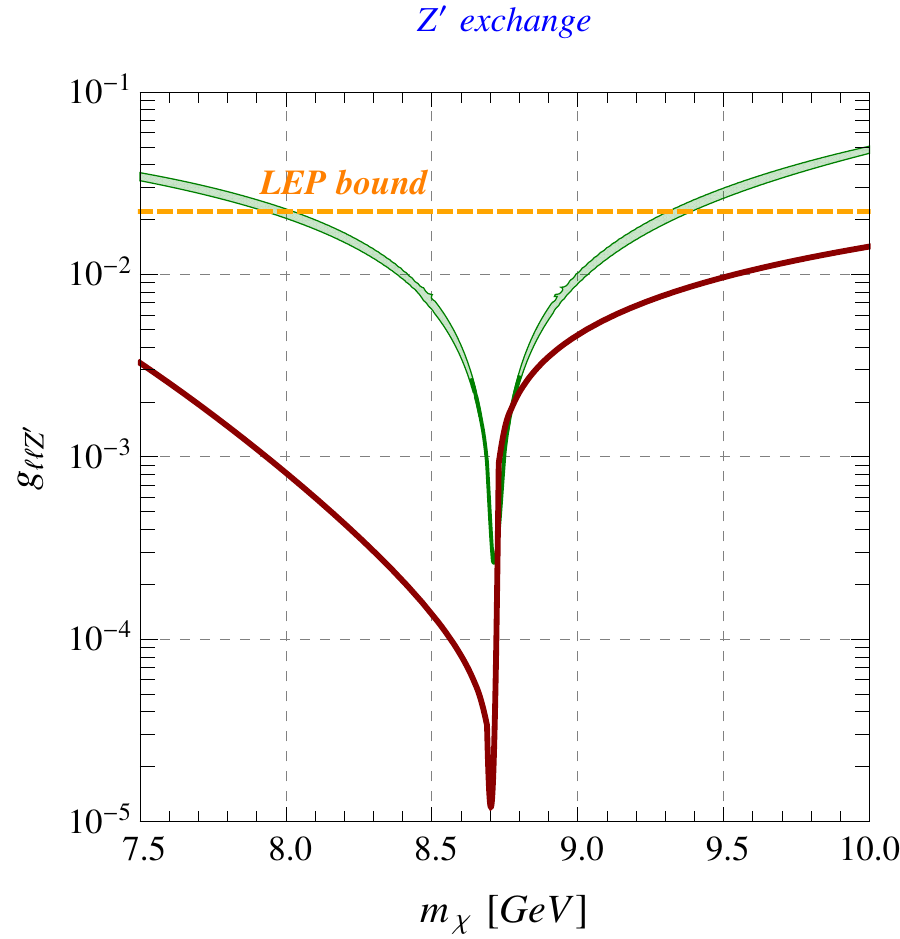}
 \caption{\emph{The $Z^\prime$-exchange case. The confidence region of the fit to the Fermi bubbles excess is in green and the right DM relic abundance is denoted by the red line. The $Z^\prime$ exchange could be the underlying mechanism for both the right DM density and the Fermi bubbles excess without violating the LEP bound, the dashed orange line.}}
 \label{fig:DM-Zprime-exchange}
 \end{figure}

Besides the Fermi bubbles excess and the DM density, we have to consider the LEP bounds.
As shown in Chapter 8 of Ref.~\cite{LEP:2003aa}, it can be cast into the cutoff $\Lambda$ in terms of contact interactions
\be
\mathcal{L} \supset \mathcal{L}_{SM} + \frac{4\pi}{ (1+\delta) \Lambda^2} \sum_{i,j=L,R} \eta_{ij}
\bar{e}_i \gamma_{\mu} e_i \bar{f}_j \gamma^{\mu} f_j \ ,
\label{eq:DM_z'_lambda}
\ee
where the constant $\delta=1(0)$ for $f=e$ $(f \neq e)$. $\eta_{ij}=1$ ($=0$) corresponds to the interactions with the positive (negative) contribution
with respect to SM ones. The main constraint comes from the large t-channel $e^+ e^- \rightarrow e^+ e^-$ contribution via
the $Z^\prime$ exchange in the limit of $m_{Z^\prime}^2 \ll s$, where $\sqrt{s}$ is the LEP center-of-mass energy, $\mathcal{O}(100)$ GeV. To obtain the bound, we compute the total cross section with SM$-Z^\prime$ interactions Eq.~(\ref{eq:DM_z'})
and compare it with the one from the contact term
Eq.~(\ref{eq:DM_z'_lambda}).

In Fig.~\ref{fig:DM-Zprime-exchange}, we show a viable example of $Z^\prime$ models, given $m_{Z^\prime} = 17.5 ~\mathrm{GeV}$.
The green area represents the $99~\%$ CL confidence region of the Fermi bubbles excess
and the red line is the relic density contour. Similar to the Higgs exchange, the overlapping region takes place near the resonance.
The LEP constraint is represented by the dashed orange line, which is well above the overlapping region between the confidence region of the fit to the Fermi bubbles excess and the relic density contour. To sum, the $Z^\prime$ exchange with only leptonic couplings could be responsible for the Fermi bubbles excess and the correct DM density.


\section{Conclusions}
\label{sec:Conclusions}
As recently revealed, the analysis of the energy spectrum of the Fermi bubbles show the
existence of a component peaked at low latitude around $E_{\gamma}\sim 1-4$ GeV.
In the previous work~\cite{Huang:2013pda} -- dedicated to the astrophysical analysis of this signal -- we argued that its origin could be compatible with DM annihilation.

In this paper, we have analyzed this excess from the point of view of particle physics, aiming at
 understanding which kind of DM and which kind of interactions are able to reproduce the distinctive features of the signal without contradict  the existing phenomenological constraints.

First, we have performed a model-independent analysis based on a two-dimensional fit using as free parameters the dark matter mass $M_{\rm DM}$ and the thermally averaged annihilation cross section $\langle \sigma v\rangle$. We have found that the signal can be accommodated
by DM annihilation into SM fermions, varying from $M_{\rm DM} \sim 10$ GeV,
 $\langle \sigma v\rangle \sim 6 \times 10^{-27}$ cm$^3$s$^{-1}$ (annihilation into $\tau^+\tau^-$) to
 $M_{\rm DM} \sim 60$ GeV, $\langle \sigma v\rangle \sim 2 \times 10^{-26}$ cm$^3$s$^{-1}$ (annihilation into $b\bar{b}$).

Second, we have investigated both fermionic and scalar DM using the language of effective operators in order to classify  all the possible structures arising at lowest order from the interaction of DM with the SM fermions. In order to scrutinize the resulting list of operators, the Occam's razor used in our analysis
relies on four basic requirements: \textit{i}) the flux of photons originating from DM annihilation must reproduce the
aforementioned gamma-ray signal, \textit{ii}) the DM relic density must be in agreement with the value observed by the Planck collaboration, \textit{iii}) the SI elastic cross section on nuclei must be consistent with the stringent bound placed by the XENON100 experiment, and \textit{iv}) the best-fit values for the DM mass and coupling must validate
the effective field theory description.

We have found that only two cases are able to fulfill all these four requirements:
fermionic DM with mass $M_{\rm DM}\simeq 52$ GeV, annihilating into $b\bar{b}$ via a vector-type interaction, and fermionic DM with mass $M_{\rm DM}\simeq 10$ GeV, annihilating into $\tau^+\tau^-$
via a pseudoscalar-type interaction. These results can serve as guiding principles for searching for UV complete theories. 

Finally, we have investigated two concrete models:
the scalar Higgs portal, and a toy model of fermionic DM in which the mediator between the dark and the visible sector is a light $Z^{\prime}$ gauge boson.

On the one hand, the scalar Higgs portal model turns out to be particularly suitable to accommodate all the features of the observed gamma-ray signal. This is due to the fact that 
when $M_{\rm DM}\simeq 60$ GeV we are close to the Higgs resonance; in this situation the dominant annihilation channel, via on-shell Higgs decay, corresponds to $b\bar{b}$, exactly as suggested by the model-independent fit performed at the beginning of our analysis. In this mass region the scalar Higgs portal
 can reproduce the
correct amount of relic density, and avoid the bounds placed by the XENON100 and the LHC experiments.

On the other one, we have shown that a light leptophillic $Z^\prime$ model with $M_{\rm DM}\sim 10$ GeV and $m_{Z^\prime}\sim 20$ GeV can also provide a good phenomenological framework in which try to accommodate
the observed signal without be in tension with the bound placed by the LEP experiment. In this case the dominant annihilation channel is required to be $\tau^+\tau^-$.

\subsection*{Acknowledgement}

We thank Marco Farina and Roberto Trotta for useful advices and
Gabrijela Zaharijas for enlightening discussions. The work of A.U. is supported by the ERC Advanced Grant n$^{\circ}$ $267985$, ``Electroweak Symmetry Breaking, Flavour and Dark Matter: One Solution for Three Mysteries" (DaMeSyFla). W.-C.H. would like to thank the hospitality of Northwestern University and IFPA at Universit\'{e} de Li\`{e}ge, where part of this work was performed.

\subsection*{Note added in proof}
After completing the paper, LUX results~\cite{Akerib:2013tjd} become available, setting a WIMP-nucleon SI cross section limit at $2 \times 10^{-46}$ cm$^2$.
Our conclusions, however, remain mostly unchanged except for the vector operator with the $b\bar{b}$ final state in the context of fermonic DM, where part of the Fermi bubbles
confidence region is in tension with the LUX results.  

\appendix

\section{Fermi bubbles energy spectrum after ICS subtraction}\label{app:0}

In this Appendix, we show the data used in this paper. The data describe the energy spectrum of the Fermi bubbles after subtracting the ICS component, from an additional population of GeV-TeV electrons trapped inside the Fermi bubbles. For the subtraction procedure, we follow the same approach adopted in Ref.~\cite{Hooper:2013rwa}. As discussed in Ref.~\cite{Huang:2013pda}, we do not use the data in the region of $|b|=1^{\circ}-10^{\circ}$, because of the large astrophysical uncertainties. We collect the energy spectrum of the Fermi bubbles after ICS subtraction, i.e. the Fermi bubbles excess analyzed throughout this paper,  in Table~\ref{tableDATA1} and Table~\ref{tableDATA2}.

In addition, the angular average of the $J$ factor used in Sec.~\ref{sec:FermiBubbles} for the generalized NFW profile is,
 \be
 \bar{J}_{\rm gNFW}=(26.49, 10.48, 5.63, 3.56),
 \ee
 as opposed to the NFW profile,
  \be
 \bar{J}_{\rm NFW}=(17.15, 8.14, 4.86, 3.00),
 \ee
 where the 4 entries correspond to 4 slices from $|b|=10^\circ - 50^\circ$, $10^o$ for each slice.


\begin{table}[!h!]
\centering
\begin{tabular}{||c|c|c||c|c||}\hline
& \multicolumn{2}{c||}{{\color{bluscuro}{$|b|=10^{\circ}-20^{\circ}$}}} &
 \multicolumn{2}{c||}{{\color{bluscuro}{$|b|=20^{\circ}-30^{\circ}$}}}   \\ [1 pt] \hline \hline
   \multirow{2}{*}{{{\color{bluscuro}{$E_{\gamma}$ [GeV]}}}} & {\color{bluscuro}{$E_{\gamma}^2
   d\Phi/dE_{\gamma}d\Omega$}} &
    {\color{bluscuro}{$\pm\delta$}} & {\color{bluscuro}{$E_{\gamma}^2
   d\Phi/dE_{\gamma}d\Omega$}} &  {\color{bluscuro}{$\pm\delta$}} \\
     &  {\color{bluscuro}{[GeVcm$^{-2}$s$^{-1}$sr$^{-1}$]}} &
     {\color{bluscuro}{[GeVcm$^{-2}$s$^{-1}$sr$^{-1}$]}} &  {\color{bluscuro}{[GeVcm$^{-2}$s$^{-1}$sr$^{-1}$]}} &  {\color{bluscuro}{[GeVcm$^{-2}$s$^{-1}$sr$^{-1}$]}} \\ \hline\hline
$0.336606$ & $3.8563\times 10^{-7}$ & $8.56511\times 10^{-8}$  & $5.84533\times 10^{-8}$ &  $6.02678\times 10^{-8}$ \\ \hline
$0.423762$ & $1.17013\times10^{-7}$ & $7.55953\times10^{-8}$  & $8.89873\times 10^{-8}$ &  $5.49673\times 10^{-8}$ \\ \hline
$0.533484$ & $-6.39891\times10^{-9}$ & $7.204\times10^{-8}$  & $1.38655\times 10^{-7}$ & $5.31509\times 10^{-8}$ \\ \hline
$0.671617$ & $9.37591\times10^{-8}$ & $7.24396\times10^{-8}$  &  $1.54946\times 10^{-7}$ & $5.35437\times 10^{-8}$ \\ \hline
$0.845516$ & $1.77333\times10^{-7}$ & $7.30135\times10^{-8}$  &  $1.34858\times 10^{-7}$ & $5.41341\times 10^{-8}$ \\ \hline
$1.06444$ & $2.76807\times10^{-7}$ & $6.01564\times10^{-8}$ & $1.55329\times 10^{-7}$ & $4.54328\times 10^{-8}$ \\ \hline
$1.34005$ & $3.06077\times10^{-7}$ & $5.99795\times10^{-8}$  &  $1.47409\times 10^{-7}$ & $4.56531\times 10^{-8}$ \\ \hline
$1.68703$ & $4.45311\times10^{-7}$ & $6.23474\times10^{-8}$ & $1.75062\times 10^{-7}$ & $4.75242\times 10^{-8}$ \\ \hline
$2.12384$ & $2.58794\times10^{-7}$ & $6.27977\times10^{-8}$ &  $1.75355\times 10^{-7}$ & $4.88555\times 10^{-8}$ \\ \hline
$2.67376$ & $4.84638\times10^{-7}$ & $6.55294\times10^{-8}$  &  $2.03788\times 10^{-7}$ & $5.0759\times 10^{-8}$ \\ \hline
$3.36606$ & $4.4664\times10^{-7}$ & $6.77863\times10^{-8}$  & $1.50084\times 10^{-7}$ & $5.24922\times 10^{-8}$ \\ \hline
$4.23762$ & $2.83971\times10^{-7}$ & $7.01118\times10^{-8}$ &  $1.22119\times 10^{-7}$ & $5.48032\times 10^{-8}$ \\ \hline
$5.33484$ & $2.85591\times10^{-7}$ & $7.36556\times10^{-8}$ &   $1.1964\times 10^{-7}$ & $5.73127\times 10^{-8}$ \\ \hline
$6.71617$ & $2.368\times10^{-7}$ & $7.63765\times10^{-8}$  & $9.69775\times 10^{-8}$ &  $5.92534\times 10^{-8}$ \\ \hline
$8.45515$ & $5.99069\times10^{-8}$ & $7.76162\times10^{-8}$  & $1.83427\times 10^{-7}$ & $6.19328\times 10^{-8}$ \\ \hline
$10.6444$ & $1.89168\times10^{-7}$ & $8.27196\times10^{-8}$ & $6.50662\times 10^{-8}$ & $6.29612\times 10^{-8}$ \\ \hline
$13.4005$ & $-3.08699\times10^{-8}$ & $8.37303\times10^{-8}$ & $7.29646\times 10^{-9}$ & $6.42694\times 10^{-8}$ \\ \hline
$16.8703$ & $1.69117\times10^{-8}$ & $8.98631\times10^{-8}$ & $8.40104\times 10^{-8}$ & $6.90306\times 10^{-8}$ \\ \hline
$21.2384$ & $-1.594\times10^{-7}$ & $9.17258\times10^{-8}$ & $6.17253\times 10^{-8}$ & $7.2138\times 10^{-8}$ \\ \hline
$26.7375$ & $2.62676\times10^{-8}$ & $1.01643\times10^{-7}$ &  $-3.50529\times 10^{-8}$ & $7.40246\times 10^{-8}$ \\ \hline
$33.6606$ & $-1.06081\times10^{-7}$ & $1.02724\times10^{-7}$ & $-1.56237\times 10^{-7}$ & $7.3966\times 10^{-8}$ \\ \hline
$42.3762$ & $3.543\times10^{-8}$ & $1.15152\times10^{-7}$  & $4.18016\times 10^{-8}$ & $8.64957\times 10^{-8}$ \\ \hline
$53.3484$ & $-2.36955\times10^{-7}$ & $1.12265\times10^{-7}$ & $-1.9607\times 10^{-7}$ & $8.22324\times 10^{-8}$ \\ \hline
$67.1617$ & $-3.4628\times10^{-7}$ & $1.11375\times10^{-7}$ & $5.16533\times 10^{-8}$ & $9.98167\times 10^{-8}$ \\ \hline
$84.5516$ & $-2.92013\times10^{-7}$ & $1.19445\times10^{-7}$  & $1.93923\times 10^{-8}$ & $1.04475\times 10^{-7}$ \\ \hline
$106.444$ & $-1.33664\times10^{-7}$ & $1.3237\times10^{-7}$ & $3.3773\times 10^{-8}$ & $1.09813\times 10^{-7}$ \\ \hline
$134.005$ & $-1.71772\times10^{-7}$ & $1.40908\times10^{-7}$  & $-1.4609\times 10^{-7}$ & $1.00974\times 10^{-7}$ \\ \hline
$168.703$ & $-4.05648\times10^{-7}$ & $1.15535\times10^{-7}$ & $-2.08629\times 10^{-8}$ & $1.17566\times 10^{-7}$ \\ \hline
$212.384$ & $-2.59493\times10^{-7}$ & $1.33816\times10^{-7}$ & $9.88512\times 10^{-9}$ & $1.2709\times 10^{-7}$ \\ \hline
$267.376$ & $-1.66617\times10^{-7}$ &  $1.51618\times10^{-7}$  & $-1.15592\times 10^{-7}$ & $1.08631\times 10^{-7}$ \\ \hline
\end{tabular}
\caption{\emph{Eenergy spectrum of the Fermi bubbles after subtraction of the ICS component in the two slices $|b|=10^{\circ}-20^{\circ}$ and $|b|=20^{\circ}-30^{\circ}$.}}
\label{tableDATA1}
\end{table}

\begin{table}[!h!]
\centering
\begin{tabular}{||c|c|c||c|c||}\hline
& \multicolumn{2}{c||}{{\color{bluscuro}{$|b|=30^{\circ}-40^{\circ}$}}} &
 \multicolumn{2}{c||}{{\color{bluscuro}{$|b|=40^{\circ}-50^{\circ}$}}}   \\ [1 pt] \hline \hline
   \multirow{2}{*}{{{\color{bluscuro}{$E_{\gamma}$ [GeV]}}}} & {\color{bluscuro}{$E_{\gamma}^2
   d\Phi/dE_{\gamma}d\Omega$}} &
    {\color{bluscuro}{$\pm\delta$}} & {\color{bluscuro}{$E_{\gamma}^2
   d\Phi/dE_{\gamma}d\Omega$}} &  {\color{bluscuro}{$\pm\delta$}} \\
     &  {\color{bluscuro}{[GeVcm$^{-2}$s$^{-1}$sr$^{-1}$]}} &
     {\color{bluscuro}{[GeVcm$^{-2}$s$^{-1}$sr$^{-1}$]}} &  {\color{bluscuro}{[GeVcm$^{-2}$s$^{-1}$sr$^{-1}$]}} &  {\color{bluscuro}{[GeVcm$^{-2}$s$^{-1}$sr$^{-1}$]}} \\ \hline\hline
$0.336606$ & $1.81409\times  10^{-8}$ & $5.03109\times  10^{-8}$ &  $-1.7298\times10^{-8}$ & $5.88503\times10^{-8}$ \\ \hline
$0.423762$ & $1.84166\times  10^{-8}$ & $4.73835\times  10^{-8}$ &  $-4.40717\times10^{-8}$ & $5.34389\times10^{-8}$ \\ \hline
$0.533484$ & $7.66864\times  10^{-10}$ &  $4.6752\times  10^{-8}$ & $-1.41948\times10^{-8}$ & $5.1282\times10^{-8}$ \\ \hline
$0.671617$ & $-9.01721\times  10^{-8}$ &  $4.74156\times  10^{-8}$ &  $4.82142\times10^{-8}$ & $5.1377\times10^{-8}$ \\ \hline
$0.845516$ & $-2.91716\times  10^{-8}$ & $4.87876\times  10^{-8}$ & $3.144\times10^{-8}$ &  $5.12871\times10^{-8}$ \\ \hline
 $1.06444$ &   $7.51726\times  10^{-9}$ & $4.18146\times  10^{-8}$ & $7.96846\times10^{-9}$ &  $4.12139\times10^{-8}$ \\ \hline
 $1.34005$ & $7.85443\times  10^{-9}$ & $4.2711\times  10^{-8}$ &  $-5.26925\times10^{-9}$ &  $4.11044\times10^{-8}$ \\ \hline
 $1.68703$ & $3.08868\times  10^{-8}$ & $4.4806\times  10^{-8}$ &  $-7.52642\times10^{-8}$ & $4.2116\times10^{-8}$ \\ \hline
 $ 2.12384$ & $2.70824\times  10^{-8}$ & $4.64359\times  10^{-8}$ &    $3.57198\times10^{-8}$ & $4.42407\times10^{-8}$ \\ \hline
  $2.67376$ & $6.65282\times  10^{-10}$ &  $4.82948\times  10^{-8}$ &   $4.59217\times10^{-8}$ & $4.64025\times10^{-8}$ \\ \hline
  $3.36606$ & $-2.46114\times  10^{-9}$ & $5.03311\times  10^{-8}$ &  $-7.31381\times10^{-9}$ & $4.83922\times10^{-8}$ \\ \hline
  $4.23762$ & $-5.04098\times  10^{-8}$ & $5.23324\times  10^{-8}$ &     $-1.8401\times10^{-8}$ & $5.14476\times10^{-8}$ \\ \hline
  $5.33484$ & $-3.10107\times  10^{-8}$ & $5.46831\times  10^{-8}$ &     $-2.42575\times10^{-8}$ & $5.48266\times10^{-8}$ \\ \hline
  $6.71617$ & $5.17692\times  10^{-9}$ & $5.67794\times  10^{-8}$ &     $1.28129\times10^{-8}$ & $5.85591\times10^{-8}$ \\ \hline
   $8.45515$ & $1.30828\times  10^{-8}$ & $5.82118\times  10^{-8}$ &  $-1.69367\times10^{-8}$ & $6.09384\times10^{-8}$ \\ \hline
    $10.6444$ & $-4.21099\times  10^{-9}$ & $5.96395\times  10^{-8}$ &     $8.13071\times10^{-8}$ & $6.64749\times10^{-8}$ \\ \hline
    $13.4005$ &  $8.60336\times  10^{-8}$ & $6.27196\times  10^{-8}$ &  $-5.3034\times10^{-8}$ & $6.62642\times10^{-8}$ \\ \hline
    $16.8703$ & $6.97485\times  10^{-8}$ & $6.53825\times  10^{-8}$ & $4.72857\times10^{-8}$ & $7.33796\times10^{-8}$ \\ \hline
     $21.2384$ & $5.12282\times  10^{-8}$ & $6.79283\times  10^{-8}$ &  $5.36482\times10^{-8}$ & $7.77766\times10^{-8}$ \\ \hline
     $26.7375$ & $-3.5633\times  10^{-8}$ & $6.91281\times  10^{-8}$ &  $8.39935\times10^{-8}$ & $8.41912\times10^{-8}$ \\ \hline
       $33.6606$ & $9.94703\times  10^{-9}$ & $7.47468\times  10^{-8}$ & $4.32264\times10^{-8}$ & $8.85267\times10^{-8}$ \\ \hline
        $42.3762$ & $-5.66737\times  10^{-8}$ & $7.76135\times  10^{-8}$ &     $-7.1643\times10^{-8}$ & $8.91189\times10^{-8}$ \\ \hline
         $53.3484$ & $-8.92775\times  10^{-8}$ & $8.14245\times  10^{-8}$ &   $-1.17304\times10^{-10}$ & $1.01232\times10^{-7}$ \\ \hline
         $67.1617$ & $-8.53756\times  10^{-8}$ & $8.74955\times  10^{-8}$ &      $4.44777\times10^{-8}$ & $1.13931\times10^{-7}$ \\ \hline
         $84.5516$ & $-3.05588\times  10^{-8}$ & $9.63193\times  10^{-8}$ &    $-1.50363\times10^{-7}$ & $1.02067\times10^{-7}$ \\ \hline
         $106.444$ & $-5.12483\times  10^{-8}$ & $9.76844\times  10^{-8}$ &           $-1.33116\times10^{-7}$ & $1.05724\times10^{-7}$ \\ \hline
         $134.005$ & $-1.74632\times  10^{-7}$ & $8.75153\times  10^{-8}$ &   $1.99941\times10^{-8}$ & $1.3237\times10^{-7}$ \\ \hline
         $168.703$ & $-5.87378\times  10^{-8}$ & $1.04816\times  10^{-7}$ &    $6.81035\times10^{-9}$ & $1.36115\times10^{-7}$ \\ \hline
         $212.384$ & $-2.92218\times  10^{-7}$ & $5.06491\times  10^{-8}$ &  $-3.49366\times10^{-8}$ & $1.34261\times10^{-7}$ \\ \hline
         $267.376$ & $-1.33076\times  10^{-7}$ & $8.9953\times  10^{-8}$ &   $-1.98532\times10^{-7}$ & $6.80737\times10^{-8}$ \\ \hline
\end{tabular}
\caption{\emph{Energy spectrum of the Fermi bubbles after subtraction of the ICS component in the two slices $|b|=30^{\circ}-40^{\circ}$ and $|b|=40^{\circ}-50^{\circ}$.}}
\label{tableDATA2}
\end{table}

\newpage
{\tiny .}

\newpage

\section{Cross sections and scattering rates}\label{app:A}

\subsection{Annihilation cross sections: Fermionic Dark Matter}

We here show for the effective operators of fermionic DM the annihilation cross sections, expanded in powers of the relative velocity $v$
\begin{eqnarray}
c_{\rm DM}^{-1}~(\sigma v)_{\rm S}&=& \frac{N_C }{16\pi M_{\chi}}\,m_f^2 v^2\sqrt{M_{\chi}^2-m_f^2}
\left[
G_{\rm S}^{2}(M_{\chi}^2-m_f^2)+G_{\rm SA}^{2}M_{\chi}^2
\right]
~,\\
c_{\rm DM}^{-1}~(\sigma v)_{\rm PS}&=& \frac{N_C \sqrt{M_{\chi}^2-m_f^2}}{4\pi M_{\chi}}\,
m_f^2\left[
G_{\rm PS}^2(M_{\chi}^2-m_f^2)+G_{\rm PSA}^2M_{\chi}^2
\right]\nonumber\\
&&
+\frac{N_C v^2 m_f^2}{32\pi M_{\chi}\sqrt{M_{\chi}^2-m_f^2}}~
\mathcal{P}_{\rm PS}(M_{\chi}^2,m_f^2)
~,\label{eq:xsecPS}\\
(\sigma v)_{\rm V}&=& \frac{N_C\sqrt{M_{\chi}^2-m_f^2}}{4\pi M_{\chi}}
\left[
G_{\rm V}^2(2M_{\chi}^2+m_f^2)+G_{\rm VA}^2(2M_{\chi}^2-2m_f^2)
\right]\nonumber\\
&&+\frac{N_C v^2}{96\pi M_{\chi}\sqrt{M_{\chi}^2-m_f^2}}~
\mathcal{P}_{\rm V}(M_{\chi}^2,m_f^2)~,
\label{eq:xsecV}\\
c_{\rm DM}^{-1}~(\sigma v)_{\rm PV}&=&\frac{N_C G_{\rm PVA}^2\sqrt{M_{\chi}^2-m_f^2}}{4\pi M_{\chi}}\,
m_f^2+\frac{N_C v^2}{96\pi M_{\chi}\sqrt{M_{\chi}^2-m_f^2}}
~\mathcal{P}_{\rm PV}(M_{\chi}^2,m_f^2)~,\label{eq:xsecPV}\\
(\sigma v)_{\rm T}&=& \frac{N_C  \sqrt{M_{\chi}^2-m_f^2}}{\pi M_{\chi}}\,m_f^2\,\left[
G_{\rm T}^{2}(M_{\chi}^2+2m_f^2)+G_{\rm TA}^{2}(M_{\chi}^2-m_f^2)
\right]
\nonumber\\&&+
 \frac{N_Cv^2m_f^2}{24\pi M_{\chi}\sqrt{M_{\chi}^2-m_f^2}}\,
~\mathcal{P}_{\rm T}(M_{\chi}^2,m_f^2)
~,\label{eq:xsecT}
\end{eqnarray}
where $c_{\rm DM}=1$ ($4$) for Dirac (Majorana) DM, the superscript $f$ in $G$ has been dropped, and we have defined the following polynomials
\begin{eqnarray}
\mathcal{P}_{\rm PS}(x,y)&=&
G_{\rm PS}^2(2x^2-xy-y^2)+G_{\rm PSA}^2(2x^2-xy)~,\\
\mathcal{P}_{\rm V}(x,y)&=&
G_{\rm V}^2(8x^2-4xy+5y^2)+G_{\rm VA}^2(8x^2+2xy-10y^2)~,\\
\mathcal{P}_{\rm PV}(x,y)&=&
4G_{\rm PV}^2\left(2x^2-xy-y^2\right)
+G_{\rm PVA}^2\left(8x^2-22xy+17y^2\right) ~,\\
\mathcal{P}_{\rm T}(x,y)&=&
 G_{\rm T}^2(4x^2-11xy+16y^2)+G_{\rm TA}^2(4x^2+7xy-11y^2)~.
\end{eqnarray}

\subsection{Annihilation cross sections: Complex Scalar Dark Matter}

 We here present for the effective operators of scalar DM the annihilation cross sections, expanded in powers of the relative velocity $v$
\bea
   ( \sigma v)_{\rm S} &=& \frac{N_C}{8 \pi} m_f^2 \sqrt{1 - \frac{m_f^2}{M_\phi^2}} \left[ F_{\rm S}^2 \left( 1- \frac{m_f^2}{M_\phi^2} \right)
      +  F_{\rm SA}^2 \right] + \frac{ N_c v^2 m_f^2 \mathcal{P}^s_{\rm S} (M_{\phi}^2,m_f^2) }{64 \pi M_\phi^4 \sqrt{ 1 - \frac{m_f^2}{M_\phi^2}} }~, \label{eq:ScalarXSectionS}
   \\
   ( \sigma v)_{\rm VS} &=& \frac{N_C}{8 \pi} m_f^2 M_\phi^4\sqrt{ 1 - \frac{m_f^2}{M_\phi^2}}
      \left[ F_{\rm VS}^2 \left( 1- \frac{m_f^2}{M_\phi^2} \right) +  F_{\rm VSA}^2 \right] + \frac{ N_c v^2 m_f^2  \mathcal{P}^s_{\rm VS} ( M_\phi^2, m_f^2) }{64 \pi \sqrt{ 1 - \frac{m_f^2}{M_\phi^2}} } ~,
      \label{eq:ScalarXSectionVS} \nonumber \\
   \\
   ( \sigma v)_{\rm V} &=& \frac{N_C}{ 24 \pi} v^2 M_\phi^2\sqrt{ 1 - \frac{m_f^2}{M_\phi^2}}
      \left[ F_{\rm V}^2 \left( 2 + 1 \frac{m_f^2}{M_\phi^2}  \right) +2 F_{\rm VA}^2 \left( 1 - \frac{m_f^2}{M_\phi^2} \right) \right]~,\label{eq:ScalarXSectionV}
   \\
   ( \sigma v)_{\rm T} &=& \frac{N_C}{ 6 \pi} v^2 m_f^2 M_\phi^4 \sqrt{ 1 - \frac{m_f^2}{M_\phi^2}}
      \left[ F_{\rm T}^2 \left( 1 + 2 \frac{ m_f^2}{M_\phi^2} \right) + F_{\rm TA}^2 \left ( 1 - \frac{m_f^2}{M_\phi^2} \right) \right]~,\label{eq:ScalarXSectionT}
\eea
where the superscript $s$ in $F$ has been dropped, and we have defined the following polynomials
\bea
   \mathcal{P}^s_{\rm S}( x,y)  &=& 3x F_{\rm S}^2 \left(x^2   - xy \right) +  F_{\rm SA}^2  xy ~,
   \\
   \mathcal{P}^s_{\rm VS}( x,y)  &=& F_{VS}^2 \left( 8x^2   - 13 xy +  5 y^2 \right) +
   F_{\rm VSA}^2 \left( 8x^2  - 7 x y\right)~.
\eea

\subsection{Direct Detection at the tree level}\label{App:TreeLevelDD}

We here provide two results of the SI DM-nucleon cross section: the scalar operator for scalar DM used in Fig.~\ref{fig:ScalarDarkMatter}, and the vector operator for fermionic DM employed to exclude the light quark final states in Section~\ref{sec:FermionicDM}.
\begin{itemize}
   \item
      {\underline{Scalar operator $\mathcal{O}_S^s$}}.
   The cross section of DM scattering off quark, $q$, is given by
   \be
      \sigma_{SI}^S= \frac{   f_q^2 F_S^2 \mu^2 m_N^2} { 8 \pi M_\phi^2 }~,
   \ee
   where $\mu = m_N M_\phi / ( m_N + M_\phi)$ is the DM-nucleon reduced mass,
   \be
     f_{c,b,t} = \frac{2}{27} \left( 1- \sum_{q= u,d,s} f_q \right),
   \ee
and
 \be
  f_u = 0.024~, \quad f_d = 0.034~, \quad f_s  = 0.046~.
 \ee
   \item
      {\underline{Vector operator $\mathcal{O}_V^f$}}.
   The cross section of DM scattering off quark, $q$, is given by
   \be
      \sigma_{SI}^f = c_{\rm DM} \frac{ \mu^2 G_V^2 } { 2 \pi  } f_{Nq}^2~,
   \ee
   where $f_{pu}=f_{nd}=2$, $f_{pd}=f_{nu}=1$, and again $c_{\rm DM}=1$ ($4$) for Dirac (Majorana) DM.
\end{itemize}

\subsection{Direct detection at one loop: the photon exchange}\label{app:DDoneLoop}
We sketch in this Appendix the details of our one loop computation for the elastic process
\begin{equation}
\chi(k)+N(p) \to \chi(k^{\prime})+N(p^{\prime})~,
\end{equation}
describing DM scattering on a nucleus at rest with charge Z and mass $m_N$ as shown in  Fig.~\ref{fig:LoopDiagram}.
The differential cross-section is given by
 \begin{equation}
 \frac{d\sigma}{dE_{\rm nr}}=\frac{\overline {|\mathcal{M}|^2}}{32\pi m_N M_{\chi}^2v^2}~,
 \end{equation}
 where $M_{\chi}$ is the DM mass, and $v$ is the DM velocity. The nuclear recoil energy is $E_{\rm nr}=E_{\chi}-E'_{\chi}$.
\begin{figure}[!htb!]
   \centering
   \includegraphics[scale=0.8]{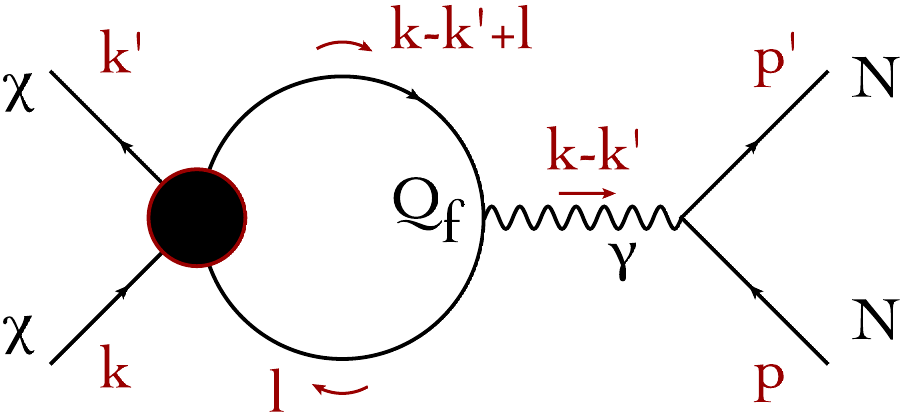}
 \caption{\emph{One-loop diagram describing the DM-Nucleus elastic scattering through the exchange of a photon.}}
 \label{fig:LoopDiagram}
\end{figure}
We here consider only the vector operator in Eq.~(\ref{eq:SMVector}). The amplitude $\mathcal{M}$ is given by the following expression
\begin{equation}\label{eq:skeleton}
i\mathcal{M}=-\frac{e^2Q_f}{\sqrt{2}}\left[
\bar{u}(k')\gamma^{\mu}u(k)
\right]\times \mathcal{I}_{\mu\sigma}\times
\langle N(p')|\sum_i Q_i \bar{q}_i\gamma^{\sigma}q_i|N(p)\rangle~,
\end{equation}
where $Q_f$ is the electric charge, in units of $e$, of the fermion running in the loop, and we sum over contributions from all the light quarks $q_i$ of electric charge $Q_i$. The loop integral is
\begin{equation}
\mathcal{I}_{\mu\sigma}=N_{C}\int\frac{d^4l}{(2\pi)^4}\frac{{\rm Tr}\left[
\gamma_{\mu}(G_{\rm V}^{f}+G_{\rm A}^f\gamma^5)(\slashed{l}+m_f)\gamma_{\sigma}
(\slashed{k}-\slashed{k}'+\slashed{l}+m_f)
\right]}
{(k-k')^2(l^2-m_f^2)[(k-k'+l)^2-m_f^2]}~.
\end{equation}
Using dimensional regularization and the $\overline{{\rm MS}}$ scheme we find
\begin{equation}
\mathcal{I}_{\mu\sigma}=-\frac{iG_{\rm V}^f g_{\mu\sigma}N_{C}}{36\pi^2 m_N E_{\rm nr}}
\left[
3(m_f^2-E_{\rm nr}m_N)B_{0}(-2m_NE_{\rm nr},m_f^2,m_f^2)-3A_0(m_f^2)+E_{\rm nr}m_N+3
m_f^2
\right]~,
\end{equation}
where in $D=4-2\epsilon$ dimensions
\begin{eqnarray}
A_0(m^2)&=& \frac{(2\pi\mu)^{2\epsilon}}{i\pi^2}\int d^Dk~ \frac{1}{(k^2-m^2)}~,\label{eq:A0}\\
B_0(p^2,m_1^2,m_2^2)&=&\frac{(2\pi\mu)^{2\epsilon}}{i\pi^2}\int d^Dk~ \frac{1}{\left[(p+k)^2-m_1^2\right](k^2-m_2^2)}~.\label{eq:B0}
\end{eqnarray}
Explicitly
\begin{eqnarray}
B_0(p^2,m_1^2,m_2^2)&=&\frac{1}{\bar{\epsilon}} + 2-\ln\frac{m_1m_2}{\mu^2}+\frac{m_1^2-m_2^2}{p^2}\ln\frac{m_2}{m_1}\nonumber\\
&& +\frac{\lambda^{1/2}(p^2,m_1^2,m_2^2)}{p^2}
\left\{
\begin{array}{cc}
  \ln\frac{m_1^2+m_2^2-p^2+\lambda^{1/2}(p^2,m_1^2,m_2^2)}{2m_1 m_2}~, & p^2\leqslant  a(m_1^2,m_2^2,p^2)   \\
    &    \\
 i\pi+ \ln\frac{-m_1^2-m_2^2+p^2-\lambda^{1/2}(p^2,m_1^2,m_2^2)}{2m_1 m_2}
 ~, & p^2 > a(m_1^2,m_2^2,p^2)~,
\end{array}
\right.\nonumber\\
\\
A_0(m^2)&=& m^2\left(\frac{1}{\bar{\epsilon}} + 1-\ln\frac{m^2}{\mu^2}\right)~,
\end{eqnarray}
with  $1/\bar{\epsilon}\equiv 1/\epsilon - \gamma_{\rm E} + \ln4\pi$, $a(m_1^2,m_2^2,p^2) \equiv m_1^2+m_2^2+\lambda^{1/2}(p^2,m_1^2,m_2^2)$, and $\lambda(x,y,z)\equiv x^2+y^2+z^2-2(xy+yz+xz)$.\\
The amplitude in Eq.~(\ref{eq:skeleton}) thus becomes
\begin{equation}\label{eq:VectorOneLoopAmp}
\mathcal{M}=\frac{e^2Q_fG_{\rm V}^{f}N_C}{36\sqrt{2}\pi^2m_NE_{\rm nr}}\left[
\bar{u}(k')\gamma^{\mu}u(k)
\right]g_{\mu\sigma}\left[
Z\mathcal{F}(q^2)\bar{u}_{N}(p')\gamma^{\sigma}u_{N}(p)
\right]\mathcal{L}~,
\end{equation}
where
\begin{equation}\label{eq:loop}
\mathcal{L} \equiv \left[
3(m_f^2-E_{\rm nr}m_N)B_{0}(-2m_NE_{\rm nr},m_f^2,m_f^2)-3A_0(m_f^2)+E_{\rm nr}m_N+3
m_f^2
\right]~.
\end{equation}
In Eq.~(\ref{eq:VectorOneLoopAmp}), we have made use of the relation $\langle N(p')|\sum_i Q_i \bar{q}_i\gamma^{\sigma}q_i|N(p)\rangle=Z\mathcal{F}(q)\bar{u}_{N}(p')\gamma^{\sigma}u_{N}(p)$, where $\mathcal{F}(q^2)$ is the charge nuclear form factor extracted from the electron-nucleus elastic cross section data, with $q^2=-2m_NE_{\rm nr}$. Throughout this work, we adopt for $\mathcal{F}(q)$ the Helm form factor \cite{Helm:1956zz}.\footnote{In the present case, the Helm form factor is particularly appropriate, given that we are considering DM elastic scattering mediated by the electromagnetic interaction. Note that in general, since the DM particles are
insensitive to the electromagnetic interaction, the
point-like matter distribution of the proton should be used instead of the charge distributions. See Ref. \cite{Co':2012ht} for a recent critical discussion.}  For the cross section, we finally obtain
\begin{equation}\label{eq:xsec}
 \frac{d\sigma}{dE_{\rm nr}}=\frac{\alpha^2 G_{\rm V}^{f,2}Q_f^2Z^2N_C^2}{2304\pi^3m_N^3M_{\chi}^2v^2
 E_{\rm nr}^2} \mathcal{F}^2(q^2)f(E_{\rm nr},v^2)
 |\mathcal{L}|^2~,
\end{equation}
where we have defined the following form factor
\begin{eqnarray}
f(E_{\rm nr},v^2)&\equiv&
32 E_{\rm nr}^2 m_N^2-32 E_{\rm nr} m_N^3-64 E_{\rm nr} m_N^2 M_{\chi}
-32 E_{\rm nr} m_N M_{\chi}^2+64 m_N^2 M_{\chi}^2\nonumber\\&&
+(64 m_N^2 M_{\chi}^2-32 E_{\rm nr} m_N^2
 M_{\chi})v^2 +16 m_N^2 M_{\chi}^2 v^4~.
\end{eqnarray}
Notice that in logarithmical approximation
\begin{equation}
B_0(-2E_{\rm nr}m_N,m_f^2,m_f^2)= \ln\frac{\mu^2}{m_f^2}~,~~~~~A_{0}(m_f^2)=m_f^2\ln\frac{\mu^2}{m_f^2}~,
\end{equation}
Eq.~(\ref{eq:loop}) becomes
\begin{equation}
|\mathcal{L}|^2 \sim 81 E_{\rm nr}^2m_N^2~,
\end{equation}
thus canceling $E_{\rm nr}^2$ in the denominator of Eq.~(\ref{eq:xsec}) to avoid the singularity in the limit of $E_{\rm nr} \rightarrow 0$.

We take the renormalization scale $\mu$ to be equal to the cut-off $\Lambda$ of the effective theory, making use of  the formal
substitution $1/\bar{\epsilon} + \ln\mu^2 = \ln\Lambda^2$.\footnote{
To be more precise, this replacement implies
that the one loop dependence on the renormalization scale $\mu$
 is cancelled by the same scale-dependence of the appropriate counterterms, which absorb the UV-divergences and
 are renormalized at the cut-off scale $\Lambda$.
In the present case the counterterm is provided by the operator $\mathcal{O}_{\rm M}
=\mathcal{C}_{\rm M}\bar{\chi}\gamma^{\mu}\chi(\partial^{\nu}F_{\mu\nu})$
whose contribution to the amplitude in Eq.~(\ref{eq:skeleton}) is given by
\begin{equation}
i\mathcal{M}_{\rm M}=-i\mathcal{C}_{\rm M}
\left[
\bar{u}(k')\gamma^{\mu}u(k)
\right]g_{\mu\sigma}
\langle N(p')|\sum_i Q_i \bar{q}_i\gamma^{\sigma}q_i|N(p)\rangle~,
\end{equation}
where $\mathcal{C}_{\rm M}=\mathcal{C}_{\rm M}^0 + \mathcal{C}_{\rm M}(\mu)$.
The cancellation of the UV-divergences in the $\overline{{\rm MS}}$ scheme implies
\begin{equation}
\mathcal{C}_{\rm M}^0 = -\frac{e^2Q_fG_{\rm V}N_{\rm C}}{12\sqrt{2}\pi^2 \bar{\epsilon}}~,
\end{equation}
while the counterterm renormalized at the scale $\Lambda$ is
\begin{equation}
\mathcal{C}_{\rm M}(\Lambda) \equiv \mathcal{C}_{\rm M}(\mu) -
\frac{e^2Q_fG_{\rm V}N_{\rm C}}{12\sqrt{2}\pi^2}\ln\frac{\Lambda^2}{\mu^2}~.
\end{equation}
In our analysis we assume $\mathcal{C}_{\rm M}(\Lambda)=0$.
}

\subsection{Scattering rates}\label{app:scatteringrates}
In this Appendix, we briefly review the basic ingredients used in the analysis of the XENON100 data.
The differential scattering rate, measured in $events\times kg^{-1}\times days^{-1}\times keV^{-1}$, is given by
\begin{equation}\label{eq:diffract}
\frac{dR}{dE_{\rm nr}}=\frac{N_{\rm T}\rho_{\odot}}{M_{\chi}}
\int_{|\vec{v}|> v_{\rm min}}d^3 \vec{v}\,|\vec{v}|\,f(\vec{v})\, \frac{d\sigma}{dE_{\rm nr}}~,
\end{equation}
where $N_{\rm T}\equiv N_{\rm Av}/A$ is the number of nuclei in the target per unit detector mass with $N_{\rm Av}=6.02\cdot10^{26}$ kg$^{-1}$ and $A$ being the mass number of the target nucleus. $f(\vec{v})$ is the DM velocity distribution, and $v_{\rm min}$ is the required minimal velocity for given nuclear recoil energy $E_{\rm nr}$. For the DM velocity distribution w.r.t. the GC, we use the Maxwellian profile
\begin{equation}\label{eq:Maxwell}
\tilde{f}(\vec{u}) = \frac{1}{N_{\rm esc}}\,e^{-|\vec{u}|^2/v_0^2}\,\vartheta(v_{\rm esc}-|\vec{u}|)~,
\end{equation}
 where $N_{\rm esc}(z)\equiv [{\rm Erf}(z)-2ze^{-z^2}\pi^{1/2}]\pi^{3/2}v_0^3$, $v_{\rm esc}=544$ km s$^{-1}$, and $v_0=220$ km s$^{-1}$. In order to obtain the DM velocity distribution seen on Earth in Eq.~(\ref{eq:diffract}), one has to include the relative velocity of the Earth w.r.t. the GC, $\vec{v}_{\rm obs}$. Therefore, we have $f(\vec{v})=\tilde{f}(\vec{v}+\vec{v}_{\rm obs})$; please see Ref.~\cite{Savage:2006qr} for more details.
 The latest results of the XENO100 experiment \cite{Aprile:2012nq}
 have been obtained analyzing $224.6$ live days $\times$  $34$ kg exposure.
 Two events have been observed in the nuclear recoil
energy range $E_{\rm nr}=6.6-30.5$ keV, consistently with the expected number of events from the background, i.e. $b=1.0\pm 0.2$.
In this paper we perform a chi-square analysis of the events, following the procedure outlined in Refs.~\cite{Farina:2011bh,Farina:2011pw} (see also Refs.~\cite{Aprile:2011hx,Bertone:2011nj}).

\section{Boltzmann equation and relic density}\label{App:Boltzmann}

The expanding of the Universe and the annihilation of DM will make DM density $n$ deviate from its equilibrium value $n_{eq}$,
which follows the Boltzmann equation is the Universe
\begin{equation}
a^{-3} \frac{\d \left(  n a^3\right)}{\d t}  =   - c \langle \sigma v \rangle \left( {n^2}-n_{eq}^2 \right) \ ,
\end{equation}
where $a$ is the scale factor of the expanding Universe.
For the real scalar and Majorana fermion, the symmetry factor $c$ is 1, and for the complex scalar and Dirac fermion, $c$ is $\frac{1}{2}$. 
We will NOT write explicitly the factor, but we include it in all the computation.

The equation can be simplified by using the variable $Y = \frac{n}{s}$ and $x=\frac{m_X}{T}$,
where $s$ is the entropy density of the Universe
\begin{equation}
 s  = \frac{2 \pi^2}{45} h_{eff} T^3 \ ,
\end{equation}
$T$ is the Universe temperature, and $h_{eff}$ is the effective entropy degrees of freedom.
The Boltzmann equation can be rewritten as
\begin{equation}
   \frac{\d Y } {  \d x}  = -\frac{ \lambda } {x^2} \left( Y^2 - Y_{eq}^2 \right) \ ,
   \label{eq_boltz}
\end{equation}
and
\begin{equation}
   \lambda = \sqrt{ \frac{ \pi} { 45}  } m_\chi M_{pl}  \left[ \sqrt{ g_\star }   < \sigma v > \right] (x)  \ .
\end{equation}
The square root of the effective degrees of freedom is defined as follows, and showed in Fig.~\ref{fig_hgeff}
\begin{equation}
   \sqrt{ g_\star} = \frac{ h_{eff} } { \sqrt{g_{eff}}  } \left( 1 + \frac{T} { 3 h_{eff} } \frac{ \d h_{eff} }{ \d T}\right) \ ,
\end{equation}
where $g_{eff}$ is the effective energy degrees of freedom in the Universe.

\begin{figure}[!htb!]
   \begin{minipage}{0.4\textwidth}
   \centering
   \includegraphics[scale=0.6]{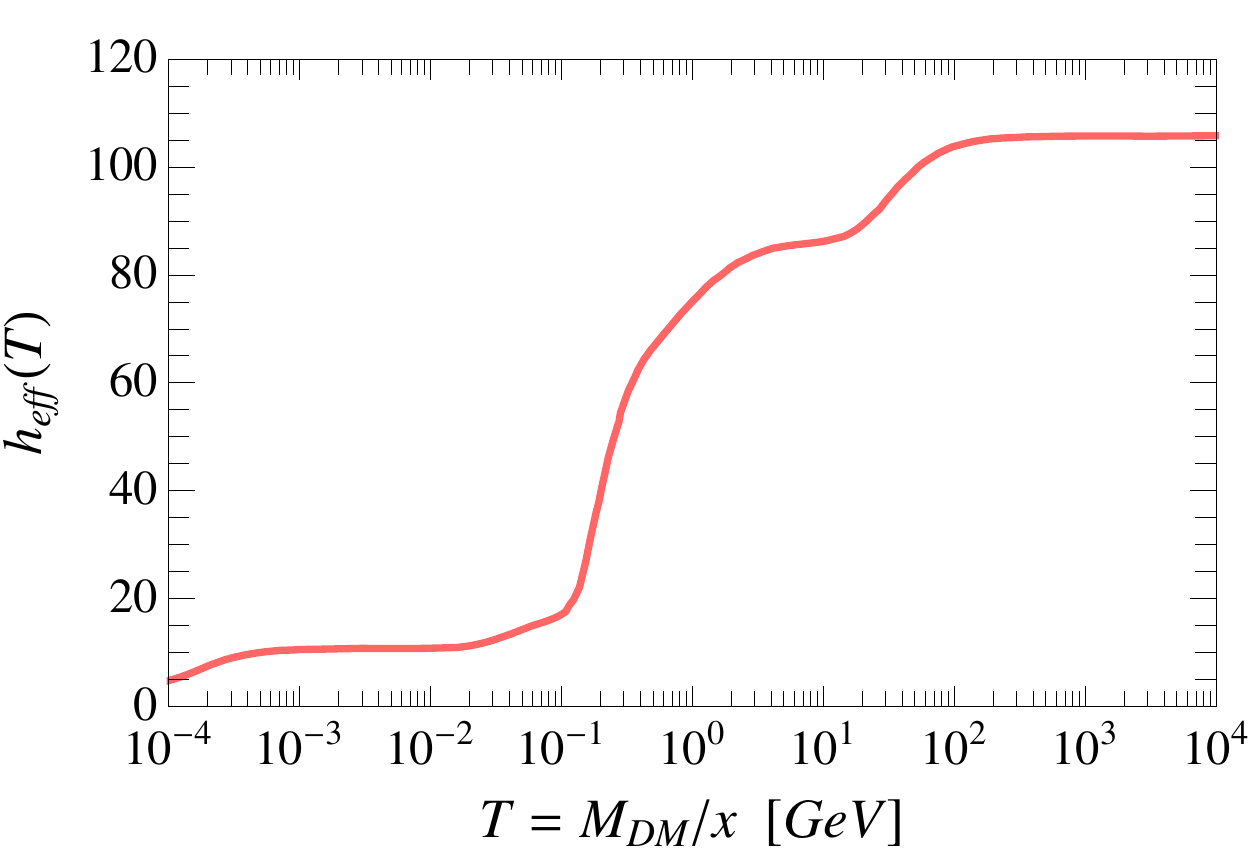}
    \end{minipage}\hspace{1 cm}
   \begin{minipage}{0.4\textwidth}
    \centering
    \includegraphics[scale=0.65]{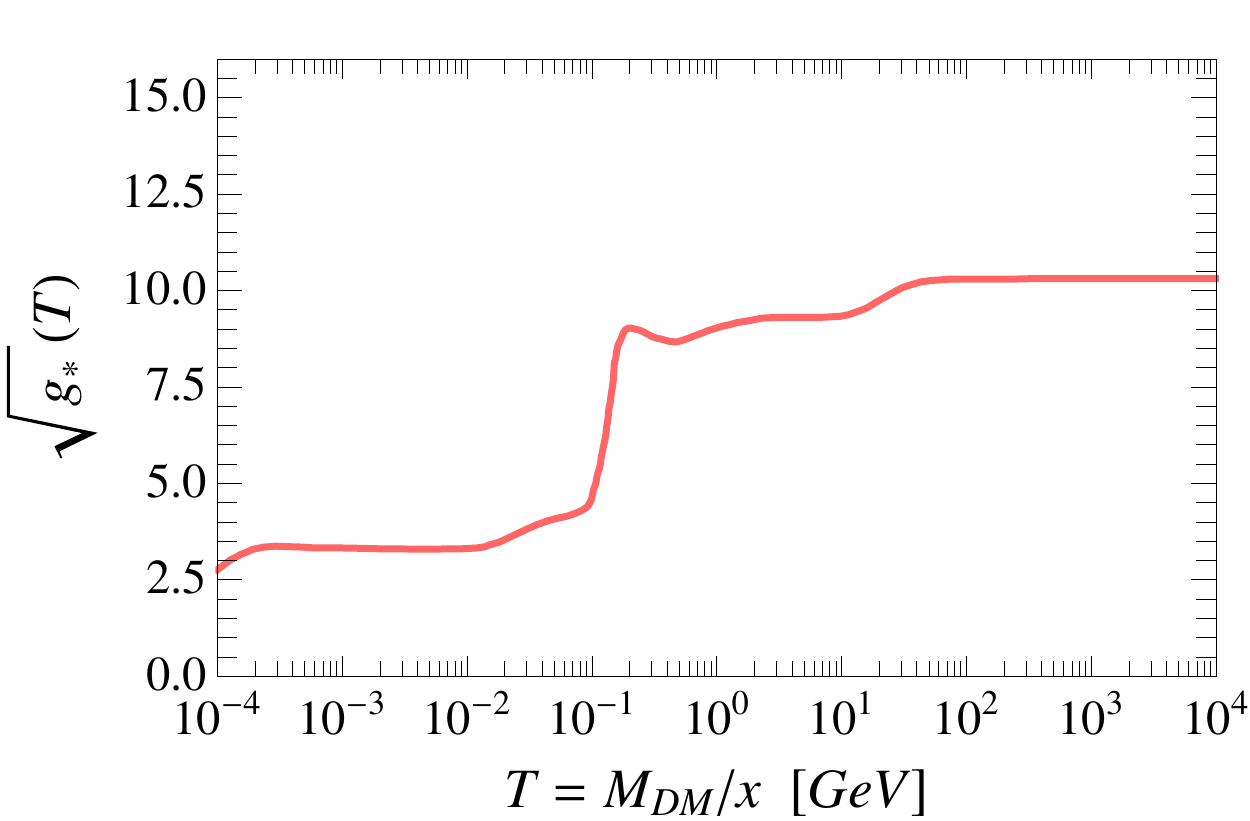}
    \end{minipage}
 \caption{\emph{The left curve the effective entropy degrees of freedom; the right curve the square root of the
effective degrees of freedom in the Universe. }}
\label{fig_hgeff}
 \end{figure}

DM initially is in the thermal equilibrium in the Universe plasma, $Y \simeq Y_{eq}$, which is the initial condition of the Boltzmann equation. 
In the Maxwell-Boltzmann approximation
\begin{equation}
\label{eq:Yin}
   Y_{eq}(x)  = \frac{45}{4 \pi^4} \frac{g ~ x^2  } {h_{eff}( m/ x) }  K_2 (x) \ ,
\end{equation}
where the factor $g$ is the internal~(spin) degree freedom of the particle. 
At late time, after DM freezes
out, its number density is much larger than its equilibrium value. Therefore, 
we can neglect $Y_{eq}$ in the Eq.~(\ref{eq_boltz}) and
integrate the equation exactly
\begin{equation}
\label{eq:Yinf}
Y_\infty  = \frac{ Y_{f} } { 1+  Y_f   \int_{ x_f}^\infty  \d x \frac{ \lambda } {x^2} } \ ,
\end{equation}
where the subscript $f$ represents the value at the freeze-out time.

To obtain the DM relic density, we can solve the Boltzmann equation numerically with its initial condition Eq.~(\ref{eq:Yin}).
Also we can solve it semi-analytically by finding the freeze-out time $x_f$ to match the initial and final value of $Y$ Eq.~(\ref{eq:Yin},\ref{eq:Yinf}).
By defining $Y = (1+ \delta ) Y_{eq}$, initially the dark matter distribution is close to $Y_{eq}$ and $\delta$ is small and grows slowly
$\d \delta / \d t \ll \delta $.
In the end $\delta$ will change dramatically, since $Y$ is close to $Y_f$. At freeze-out time $\delta \sim \mathcal{O} (1)$ is required and the
condition of $ \d \delta  / \d t \ll \delta$  still holds. Neglecting the derivative term of $\delta$, the freeze-out $x_f$ can be solved by iteration
of the Boltzmann equation
\begin{equation}
   \frac{\d \delta } {\d x} + ( 1+ \delta )  \frac{ \d \log Y_{eq} } { \d x} = - \frac{\lambda}{ x^2}Y_{eq} \delta ( 2+ \delta )
\end{equation}
 with $\delta$ chosen to a value of $\mathcal{O}(1)$.
Having the value of $Y$ at present, the reduced relic density can be written as
\begin{equation}
   \Omega h^2 \simeq 2.74 \times 10^8  ~ \frac{m_\chi}{\GeV} ~ Y_\infty  \ .
\end{equation}


\end{toexclude}

\newpage

\bibliography{Fermi_Bubbles}
\bibliographystyle{h-physrev}

\end{document}